\documentclass[%
 twocolumn, 
superscriptaddress,
 amsmath,amssymb,
 aps, %prl,
prb,
]{revtex4-2}
\usepackage{amsmath,amssymb}
\usepackage{physics}
\usepackage[dvipdfmx]{graphicx}% Include figure files
\usepackage{subfigmat}
\usepackage{xparse}
\usepackage{xstring}
\usepackage{dcolumn}% Align table columns on decimal point
\usepackage{bm}% bold math
\usepackage{mathtools}
\usepackage{xcolor}
\usepackage{hyperref}

\usepackage[normalem]{ulem}
\hypersetup{colorlinks=true,breaklinks,linkcolor=blue,urlcolor=blue,citecolor=blue}

\begin{document}

\title{Orbital Accumulation Induced by Chiral Phonons}

\author{Tetsuya Sato}
\affiliation{Institute for Solid State Physics, University of Tokyo, Kashiwa, 277-8581, Japan}

\author{Takeo Kato}
\affiliation{Institute for Solid State Physics, University of Tokyo, Kashiwa, 277-8581, Japan}

\author{Aurelien Manchon}
\affiliation{Aix-Marseille Univ, CNRS, CINaM, Marseille, France}

\date{\today}

\begin{abstract}
We theoretically investigate orbital accumulation driven by chiral phonons via orbital-dependent electron-lattice coupling. 
We derive a formula for the orbital accumulation induced by classical lattice dynamics or nonequilibrium phonons, emphasizing the rectified second-order response of the orbital moment to lattice displacement.
We show that chiral phonons primarily couple to orbital quadrupole moments and that static orbital dipole accumulation can be generated at second order in the lattice displacement. Our study provides a useful method for generating orbital accumulation without using spin-orbit interactions and suggests a strategy to boost its magnitude by harnessing band structure hot spots associated with orbital degeneracy.
\end{abstract}

\maketitle

{\it Introduction.---} 
Chiral phonons~\cite{Juraschek25}, which are lattice vibrations characterized by circular or helical atomic motion~\footnote{In our study, we define chiral phonons as lattice vibrations lacking improper rotational symmetry. See also Ref.~\cite{Juraschek25}.}, have drawn continuous interest over the past decade~\cite{Zhang2014,Zhang2015,zhang2025newadvancesphononsband,zhang2025comprehensivestudyphononchirality}. 
They are natural eigenmodes of non-centrosymmetric crystals, valley‐contrasting materials, and chiral lattices~\cite{Bozovic1984,Vonsovskii1962,McLellan1988,Zhu2018,Ishito2023a}. They can be generated coherently (e.g., by ultrafast laser pulses or piezoelectric transducers) or thermally (e.g., via temperature gradients) and are increasingly seen as potential mediators coupling lattice, charge, and spin degrees of freedom~\cite{Hamada2020,Ruckriegel2020,Sasaki2021,Yao2022,Fransson2023,Kim2023,Yao2024,Yao2024b,Ohe2024,Funato2024,Li2024,Shabala2024,Nishimura2025,Qin2025,Yokoyama2025}. 
For instance, it has been observed experimentally that thermally driven phonons in a chiral insulator can generate spins in adjacent metals, the direction of which depends on the chirality of the insulator~\cite{Ohe2024}.

Interestingly, chiral phonons are not only important in chiral crystals (e.g., $\alpha$-quartz~\cite{Ueda2023,Oishi2024}, $\alpha$-HgS~\cite{Ishito2023}, or Te~\cite{Ishito2023a,Zhang2024}), but they can also lead to important phenomena in conventional centrosymmetric systems as a reservoir of orbital angular momentum. In a recent experiment, it was demonstrated that laser-induced demagnetization is accompanied by the onset of chiral phonon excitations~\cite{Tauchert2022}, demonstrating that the transfer of angular momentum from the magnon to the phonon baths takes place on the subpicosecond scale. In the presence of spin-orbit coupling, these phonons can also polarize itinerant electrons, leading to nonequilibrium spin densities~\cite{Yao2025}. A remarkable feature of such mechanisms is the possibility of generating a spin signal without involving magnetism. Nonetheless, their efficiency is hindered by the need for large spin-orbit coupling, limiting these effects to high-Z materials. 

Chiral phonons are particularly instrumental in the context of orbitronics, an emerging field of research that aims to control the orbital angular momentum of itinerant electrons~\cite{Go2018,Lee2021b,Gao2025}. In the absence of spin-orbit coupling, the angular momentum conservation law suggests that chiral phononic currents can generate electron-mediated orbital currents. This mechanism is particularly appealing because phonons with widely differing spectral and temporal characteristics—thermal, optical, or even surface-acoustic modes—can be selectively generated and used to pump orbital currents.
For instance, a recent experiment reported that acoustic phonons generate such orbital currents~\cite{Taniguchi2025,Seungyun2025}, although this classical-displacement-based mechanism is not applicable to incoherent phonon populations with a vanishing expectation value of the lattice displacement.
Understanding the interplay between chiral phonons and orbital currents could unlock a range of experimentally relevant phenomena, from adiabatic orbital pumping to the generation of sub-picosecond orbital currents driven by nonequilibrium phonons.

In this Letter, we formulate the static orbital accumulation induced by chiral phonons without using either magnetism or spin-orbit coupling. Using a generic tight-binding model, we show that chiral phonons couple primarily to the electron’s orbital quadrupole moment, and a rectified orbital dipole moment accumulation arises at the second order in the lattice displacement. The efficiency of this effect is comparable to that of light-induced orbital moment generation, and it is expected to be greatly enhanced at band-structure hot spots featuring orbital degeneracy. 

\begin{figure}[tb]
\centering
\includegraphics[width=0.8\linewidth]{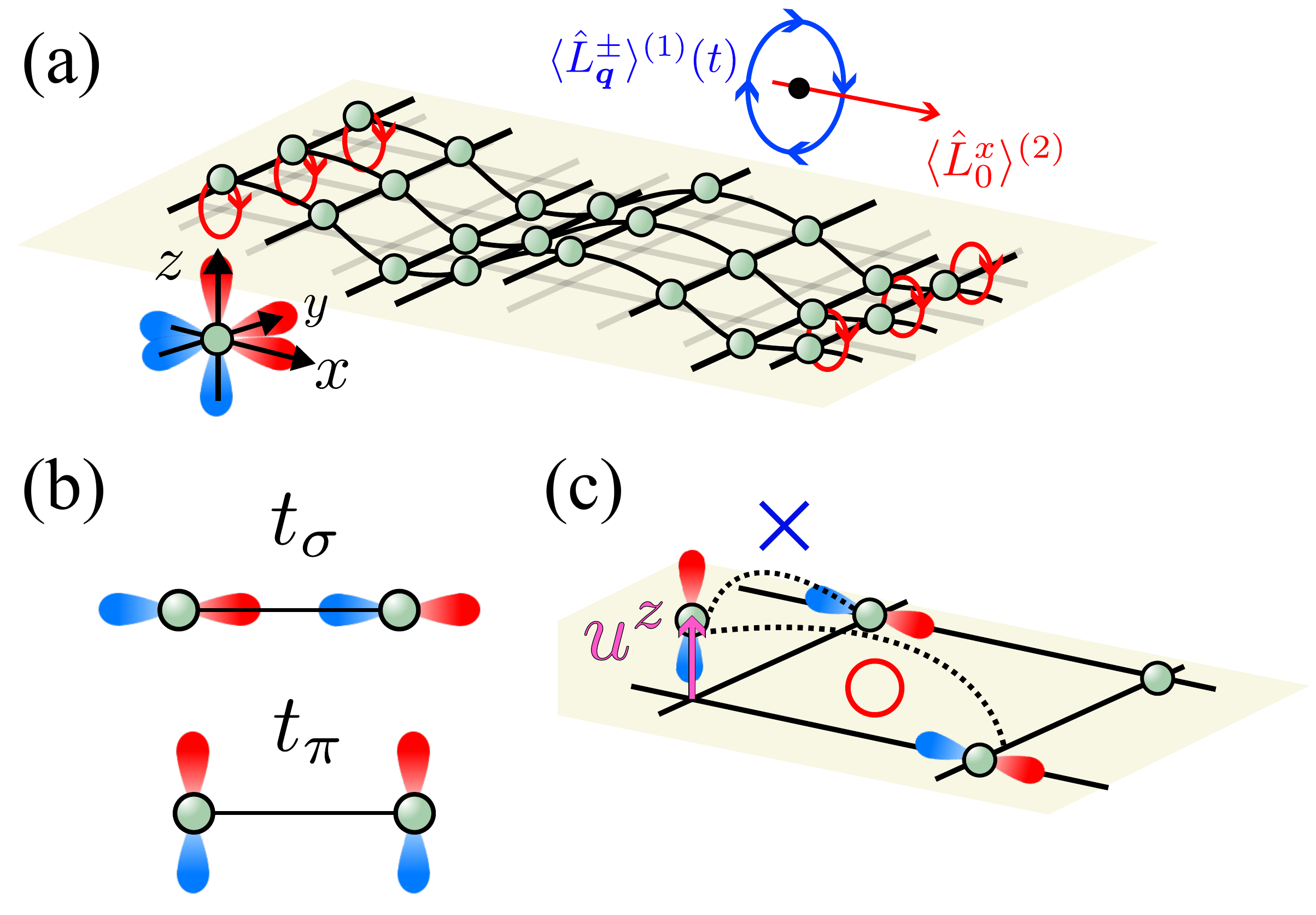}
\caption{(a) A square lattice with $p$ orbitals is embedded in the $xy$-plane, and a chiral phonon mode is injected along $+x$-axis.
(b) The schematic of the hopping integrals, $t_\sigma$ and $t_\pi$, is depicted.
(c) The displacement along $z$-axis, $u^z$, activates the hopping between $p_x$ and $p_z$ orbitals.
}
\label{fig:model}
\end{figure}

{\it Model.---} 
We consider a tight-binding model of p-orbitals on a square lattice in the $xy$ plane (Fig.~\ref{fig:model}(a)).
In the two-center hopping approximation, the tight-binding Hamiltonian is given as
\begin{align}
\mathcal{H}_0 &= \sum_{{\bm r},\alpha,\beta} \sum_{{\bm d}} t_{\alpha\beta}({\bm d}) c^\dag_{{\bm r}+{\bm d},\alpha} c_{{\bm r},\beta} , 
\end{align}
where ${\bm d}$ is a displacement vector to the nearest-neighbor sites, ${c}_{{\bm r},\alpha}$ is an annihilation operator of conduction electrons at the site ${\bm r}$, and $\alpha$ ($=x,y,z$) denotes the orbitals $p_x$, $p_y$, and $p_z$. 
Using the symmetry, the hopping parameter, $t_{\alpha\beta}({\bm d})$, is reduced to $t_\sigma$, $t_\pi$, or zero, depending on the configuration (see Fig.~\ref{fig:model}(b)).
By using the Fourier transformation, $c_{{\bm r},\alpha} = N^{-1/2} \sum_{\bm k} e^{i{\bm k}\cdot {\bm r}} c_{{\bm k},\alpha}$ ($N$: the number of sites), the Hamiltonian is diagonalized as $\mathcal{H}_0 = \sum_{\bm k,\alpha} \epsilon_{{\bm k}}^\alpha c^\dag_{{\bm k},\alpha} c_{{\bm k},\alpha}$.
The energy dispersion $\epsilon_{{\bm k}}^\alpha$ is given as
$\epsilon_{{\bm k}}^x = 2t_{\sigma}\cos k_x d  + 2t_\pi \cos k_y d-\mu$, $\epsilon_{{\bm k}}^y =2t_{\pi} \cos k_x d  + 2t_{\sigma}\cos k_y d-\mu$, $\epsilon_{{\bm k}}^z =
2t_{\pi} (\cos k_x d + \cos k_y d)-\mu$, where $\mu$ is a chemical potential, and $d$ is a lattice constant.

By taking into account the modulation of the hopping parameter up to the first order of the lattice displacement $\bm u({\bm r})$, the Hamiltonian of the electron-phonon interaction is given as
\begin{align}
\mathcal{H}_{\rm ep} 
= \sum_{\alpha\beta\bm k\bm q} {\bm \Lambda}_{\alpha\beta} ({\bm k},{\bm q}) \cdot {\bm u}_{\bm q} c^\dag_{\bm k+\bm q\alpha} c_{\bm k\beta} .
\end{align}
We first discuss orbital accumulation induced by chiral phonons using second-order perturbation with respect to $\mathcal{H}_{\rm ep}$.
For this purpose, we can drop the diagonal terms with ${\bm \Lambda}_{\alpha\alpha}$ because they do not contribute to the orbital accumulation.
Furthermore, within the two-center approximation, the coupling strength ${\bm \Lambda}_{\alpha\beta}({\bm k},{\bm q})$ can be rewritten in terms of the hopping parameters~\cite{Mitra1969,Barisic1972,Seungyun2025}.
As a result, the Hamiltonian for the orbital-dependent electron-phonon coupling is simplified as~\cite{Supplement}
\begin{align}
\mathcal{H}_{\rm ep} &= \sum_{\bm k, \bm q} \frac{2i(t_\sigma-t_\pi)}{dN} 
\\\nonumber&\times\left\{\Bigg[ \kappa_{\bm k,\bm q}^y u^z_{\bm q} c^\dag_{\bm k+\bm q,y}  c_{\bm k,z} + \kappa_{\bm k,\bm q}^x u^z_{\bm q}  c^\dag_{\bm k+\bm q,z} c_{\bm k,x} \right.
\\\nonumber&\left.\hspace{5mm}+\left(\kappa_{\bm k,\bm q}^x u^y_{\bm q} + \kappa_{\bm k,\bm q}^y u^x_{\bm q} \right) c^\dag_{\bm k+\bm q,x} c_{\bm k,y} \Bigg] + h.c. \right\},
\end{align}
where $\kappa_{\bm k,\bm q}^\mu = \sin k_\mu d - \sin (k_\mu + q_\mu) d$ ($\mu=x,y,z$). 
This Hamiltonian activates the transition between different orbitals, as shown in Fig.~\ref{fig:model}(c), and the above Hamiltonian can serve as a seed for generating the orbital moment. 

Hereafter, we consider the lattice displacement with wavenumber $\pm \bm q$ propagating along the $+x$-direction, which is described as
\begin{align}
    \bm u_i(t)&=  U_{\bm q} [\cos(-\omega_{\bm q}t+\bm q\cdot \bm R_i)\bm e_y \\
    &\quad + \lambda \sin(-\omega_{\bm q}t+\bm q\cdot \bm R_i)\bm e_z ] ,
\end{align}
where $U_{\bm q}$ is the amplitude of the displacement, $\omega_{\bm q}=v|{\bm q}|$ is the phonon dispersion, $v$ is the sound velocity, and $\lambda = \pm 1$ denotes the chirality of the phonon.
When we assume the long wavelength limit, $|\bm k|,|\bm q|\ll 1/d$ and $q_y, q_z \ll q_x$, we obtain
\begin{align}
\mathcal{H}_{\rm ep} &\simeq   \sum_{\bm q} 2C_{\bm q} \left[u_{\bm q}^y Q_{-\bm q}^{xy} + u_{\bm q}^z Q_{-\bm q}^{zx}\right]   \nonumber\\&\label{eq:elph} =
\sum_{\bm q} C_{\bm q} \left[ u_{\bm q}^+   Q_{-\bm q}^-  + u_{\bm q}^-  Q_{- \bm q}^+  \right],
\end{align}
where $C_{\bm q}=-i(t_\sigma-t_\pi)q_x/N$, $u_{\bm q}^\pm = u_{\bm q}^y\pm i u_{\bm q}^z$, $c_{\bm k,\pm}=\mp\left( c_{\bm k,y}\mp i c_{\bm k,z}\right)/\sqrt{2}$, $Q_{\bm q}^{xy}=-\sum_{\bm k}( c_{\bm k,y}^\dag c_{\bm k+\bm q,x} +  c_{\bm k,x}^\dag  c_{\bm k+\bm q,y})$, $Q_{\bm q}^{zx}=-\sum_{\bm k}( c_{\bm k,z}^\dag c_{\bm k+\bm q,x} +  c_{\bm k,x}^\dag  c_{\bm k+\bm q,z})$ and $Q_{\bm q}^\pm=Q_{\bm q}^{xy}\pm i Q_{\bm q}^{zx}$. 
We immediately see that the chiral phonons, $u_{\bm q}^\pm$, couple to the electron's quadrupole, $Q_{\bm q}^\pm$, a central feature of phonon-driven orbital generation~\footnote{Using the quadrupole operators for the orbital angular momentum of $p$-orbitals, which are defined by $Q^{x^2-y^2}=L_x^2-L_y^2$, $Q^{3z^2-r^2}= (3\hat L_z^2 - 2)/\sqrt{3}$, $Q^{xy}=L_x L_y+ L_y L_x$, $Q^{yz}=L_y L_z+ L_z L_y$, and $Q^{zx}=L_z L_x+ L_x L_z$, we obtain  $Q_{\bm q}^{\pm} = Q_{\bm q}^{xy}\pm i Q_{\bm q}^{zx}$.}.

\begin{figure}[tbp]
    \centering
    \includegraphics[width=0.95\linewidth]{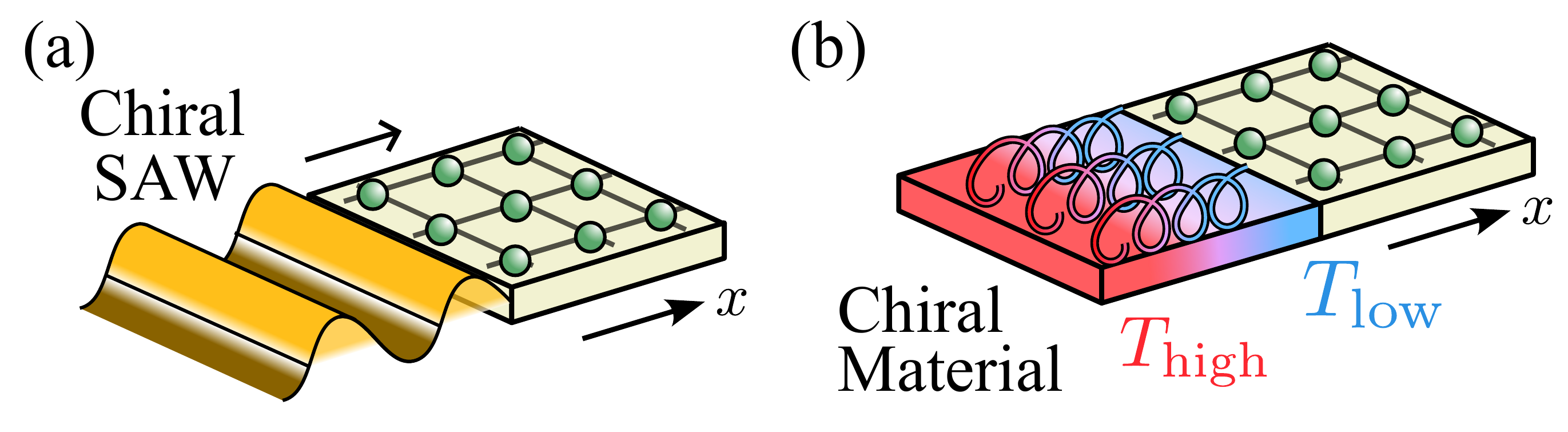}
    \caption{
    (a) Schematic illustration of the mechanism for generating a coherent chiral displacement mode using, for example, a surface acoustic wave (SAW).  
    (b) Schematic illustration of nonequilibrium thermal chiral phonon injection from an attached chiral material with a temperature gradient into the square lattice.}
    \label{fig:setup}
\end{figure}

{\it Coherent phonons and orbital moments.---}
Let us first consider the low-frequency chiral phonon induced by an external source using, e.g., a surface acoustic wave (SAW)~\cite{Liao2024,Su2025}, as shown in Fig.~\ref{fig:setup}(a).
In this case, we can assume that the lattice displacement ${\bm u}_{\bm q}$ is a classical variable.
Then, the expectation value of the orbital moment, $\langle{\bm L}_{\bm q}\rangle$, can be expressed in terms of the Berry curvature by extending the method of Ref.~\cite{Yao2025} to the orbital angular momentum.
To simplify the calculation, we introduce an auxiliary Zeeman field by the Hamiltonian $\mathcal{H}_{\rm Z}=\gamma \sum_{\bm q} \bm B_{\bm q} \cdot {\bm L}_{-{\bm q}}$ and express $\langle{\bm L}_{-{\bm q}}\rangle$ in terms of the parameter derivative of the Hamiltonian $\mathcal{H} = \mathcal{H}_0 + \mathcal{H}_{\rm ep} + \mathcal{H}_{\rm Z}$ as
\begin{align}
\langle {\bm L}_{\bm q} \rangle(t) = \frac{1}{\gamma}\Tr\left\{ \rho(t) (\partial_{\bm B_{-\bm q}} \mathcal{H})\right\}_{\bm B_{-\bm q} \rightarrow 0}.
\end{align}
The eigenstate and eigenenergy in the presence of lattice displacement and the Zeeman field are expressed up to first-order perturbation as
\begin{align}
\ket{\psi_\alpha} &\simeq \ket{\bm k \alpha}  + \sum_{\bm k' \beta} \frac{\bra{\bm k' \beta}(\mathcal{H}_{\rm ep}+\mathcal{H}_{\rm Z})\ket{\bm k \alpha}}{\epsilon_{\bm k}^\alpha-\epsilon_{\bm k'}^\beta} \ket{\bm k' \beta}, \\
\epsilon_\alpha &\simeq \epsilon_{{\bm k}}^\alpha + \bra{\bm k \alpha}(\mathcal{H}_{\rm ep}+\mathcal{H}_{\rm Z}) \ket{\bm k \alpha} .
\end{align}
Thus, up to the first order of $\dot{u}^\lambda_{\bm q}$, the density matrix for $\alpha \neq \beta$ is calculated as
\begin{align}
&\bra{\psi_\beta}\rho(t)\ket{\psi_\alpha} \nonumber \\  
& \simeq i\hbar \sum_{\bm q} \dot{u}_{\bm q}^\lambda \frac{f_\alpha  \bra{\psi_\beta}\ket{\partial_{u^\lambda_{\bm q}}\psi_\alpha} + f_\beta \bra{\partial_{u^\lambda_{\bm q}}\psi_\beta}\ket{\psi_\alpha}}{\epsilon_{\bm k}^\beta-\epsilon_{\bm k}^\alpha},
\end{align}
where $f_\alpha = (e^{(\epsilon_{\alpha}-\epsilon_{\rm F})/k_{\rm B}T}+1)^{-1}$ ($\epsilon_{\rm F}$: the Fermi energy) is the Fermi distribution function~\footnote{The density operator also includes a correction due to the lattice displacement as $\rho(t) = \rho_{\rm eq}(t) + \delta\rho(t)$, where $\bra{\psi_\beta}\delta\rho(t)\ket{\psi_\alpha} = \sum_{\bm q} \dot{\bm u}_{\bm q} \cdot \bra{\psi_\beta}i\hbar\partial_{\bm u_{\bm q}}\rho(t)\ket{\psi_\alpha}/(\epsilon_\beta - \epsilon_\alpha)$, which is estimated as $\hbar \omega_{\bm q} \langle \rho_{eq} \rangle/(\epsilon_{{\bm k}}^\beta - \epsilon_{{\bm k}}^\alpha)$. Assuming that the phonon energy $\hbar \omega_{\bm q}$ is much smaller than the energy splitting $\epsilon_{{\bm k}\beta} - \epsilon_{{\bm k}\alpha}$, we can drop this correction.}.
The orbital moment is evaluated up to second order of the lattice displacement as $\langle {\bm L}_{\bm q} \rangle (t) = \langle {\bm L}_{\bm q} \rangle_{(1)} (t) + \langle {\bm L}_{\bm q} \rangle_{(2)} (t) + {\cal O}(|{\bm u}|^3)$.
The linear term $\langle {\bm L}_{\bm q} \rangle_{(1)} (t)$, which oscillates with a phonon frequency $\omega$, is shown to be equal to the result of the linear response theory.
On the other hand, the second-order term $\langle {\bm L}_{\bm q} \rangle_{(2)} (t)$ includes a dc component, which is the central target in our study~\footnote{In principle, the second-order terms of $\langle L^\pm_{\bm q}\rangle$, which correspond to electron-hole excitation due to phonons, can exist. This effect, which is effective only near the Fermi surface, is expected to be much smaller than the Fermi sea effect discussed in the main text, assuming that the Fermi energy is much larger than the phonon energy.}. For detailed calculations, see Supplemental Material~\cite{Supplement}.

From the Hamiltonian \eqref{eq:elph}, the dc orbital accumulation is calculated for $|{\bm q}|\ll |{\bm k}|$ as
\begin{align}
& \langle L^x_{\bm 0} \rangle_{(2)} \simeq \sum_{\bm k} 4\lambda\hbar\omega_{\bm q} |C_{\bm q}|^2 U_{\bm q}^2 {\cal A}_{\bm k}, \label{mainresult} \\
& {\cal A}_{\bm k} = \frac{f_x}{(\epsilon_{\bm k}^x-\epsilon_{\bm k}^y)(\epsilon_{\bm k}^x-\epsilon_{\bm k}^z)^2} + \frac{f_x}{(\epsilon_{\bm k}^x-\epsilon_{\bm k}^z)(\epsilon_{\bm k}^x-\epsilon_{\bm k}^y)^2}  \nonumber
\\&\hspace{5mm} - \frac{f_y}{(\epsilon_{\bm k}^y-\epsilon_{\bm k}^z)(\epsilon_{\bm k}^y-\epsilon_{\bm k}^x)^2} - \frac{f_z}{(\epsilon_{\bm k}^z-\epsilon_{\bm k}^y)(\epsilon_{\bm k}^z-\epsilon_{\bm k}^x)^2} \label{calAk}.
\end{align}
The accumulation changes its sign depending on the chirality, $\lambda$, of the lattice displacement.
While orbital accumulation vanishes for $f_x=f_y=f_z$, it is enhanced by near-degeneracies between different orbital bands through the energy denominator in Eq.~\eqref{calAk} (see also the discussion in a later section).
Near such degeneracies, Eq.~\eqref{calAk} is evaluated with a finite broadening factor $\Gamma$, and the quantitative magnitude near the cusp depends on this regularization, as discussed in the Supplemental Material~\cite{Supplement}.

{\it Nonequilibrium phonons.---}
Next, we consider the nonequilibrium phonon flow induced by an adjacent chiral material, as shown in Fig.~\ref{fig:setup}(b).
In this case, the lattice displacement must be described by phonon operators as $u_{\bm q}^\pm =  (\hbar/M N\omega_{\bm q})^{1/2} (\hat a_{\bm q,\pm} + \hat a_{-\bm q,\mp}^\dag)$,
where $\hat a_{\bm q,\pm} = (\hat a_{\bm q,y} \pm i \hat a_{\bm q,z})/\sqrt{2}$.
While the orbital accumulation induced by the nonequilibrium distribution of phonons is rigorously calculated in the framework of the Keldysh formalism~\cite{Supplement}, the result is easily obtained for $|\bm q|\ll |\bm k|$ by replacing the lattice displacement as
\begin{align}
\lambda\omega_{\bm q}U_{\bm q}^2 = \frac{i\lambda}{2}\left(\dot{u}_{\bm q}^\lambda u_{-\bm q}^{\bar\lambda} +  u_{-\bm q}^{\bar\lambda} \dot{u}_{\bm q}^\lambda \right) 
\rightarrow \frac{\hbar ( n_{\bm q}^+ - n_{-\bm q}^- )}{MN},
\end{align}
and by summing it up with respect to ${\bm q}$, where $M$ is the atomic mass and $n_{\bm q}^\pm = \langle \hat a_{\bm q,\pm}^\dag \hat a_{\bm q,\pm} \rangle$ is the phonon distribution function.
Then, the orbital accumulation by nonequilibrium phonons becomes
\begin{align}
\label{eq:Lxt}
\langle L^x_0 \rangle_{(2)} &\simeq \sum_{\bm k, \bm q}  2|C'_{\bm q}|^2 (n_{\bm q}^+ - n_{\bm q}^-) \hbar\omega_{\bm q} {\cal A}_{\bm k},
\end{align}
where $C'_{\bm q}=-\sqrt{2\hbar/M N\omega_{\bm q}} C_{\bm q}$.
This expression shows that the orbital accumulation is generated by the imbalance of the chiral phonons, $n_{\bm q}^+ - n_{\bm q}^-$.

{\it Mechanism of orbital accumulation.---} 
The dc orbital accumulation can be understood as a rectification of the oscillatory orbital motion induced by the quadrupolar electron-phonon coupling. In equilibrium, the orbital wavefunctions can be chosen real and carry no orbital angular momentum. The lattice displacement generates complex interorbital superpositions through Eq.~\eqref{eq:elph}: $u_z Q^{zx}$ mixes $p_z$ and $p_x$ orbitals, whereas $u_y Q^{xy}$ mixes $p_x$ and $p_y$ orbitals. When these two quadrupolar couplings act successively, the oscillatory $p_z$-$p_x$ and $p_x$-$p_y$ motions are converted into a $p_z$-$p_y$ coherence, corresponding to $L^x$. Because the two displacement components of a chiral phonon are phase shifted by $\pi/2$, this second-order process yields the dc chirality factor $u_y\dot u_z-u_z\dot u_y \propto \lambda\omega_{\bm q}U_{\bm q}^2$. Thus, the static accumulation is a rectified second-order response whose sign is controlled by the phonon chirality.

This mechanism can be checked directly by the equations of motion~\cite{Han2022}. Up to second order in the coherent lattice displacement, we obtain
\begin{align}
&\hbar\frac{d \langle L_{\bm q,\bm k}^y \rangle }{dt} = (\epsilon_{\bm k}^x-\epsilon_{\bm k}^z) \langle Q_{\bm q,{\bm k}}^{zx} \rangle - 2i \lambda C_{\bm q}u_{\bm q}^\lambda (f_{\bm k}^z - f_{\bm k}^x ), \label{eq:Ly} \\
&\hbar \frac{d \langle L_{\bm q,\bm k}^z \rangle }{dt} =  (\epsilon_{\bm k}^y-\epsilon_{\bm k}^x) \langle Q_{\bm q,\bm k}^{xy} \rangle - 2 C_{\bm q}u_{\bm q}^\lambda (f_{\bm k}^y - f_{\bm k}^x) , \label{eq:Lz} \\
&\hbar\frac{d \langle  Q_{\bm 0,\bm k}^{yz} \rangle}{dt} = - (\epsilon_{\bm k}^z-\epsilon_{\bm k}^y)\langle  L_{\bm 0,\bm k}^{x} \rangle 
\nonumber\\&\hspace{1cm}  + \sum_{\bm q'} C_{\bm q'} \Bigg\{ u_{\bm q'}^+ \langle  L^-_{-\bm q',\bm k} \rangle - u_{\bm q'}^- \langle  L^+_{-\bm q',\bm k} \rangle \Bigg\},
\label{eq:Q_yz}
\end{align}
where $L_{\bm q}^\alpha = \sum_{\bm k} L_{\bm q,\bm k}^\alpha$ and $Q_{\bm q}^{\alpha\beta} = \sum_{\bm k} Q_{\bm q,\bm k}^{\alpha\beta}$.
The first two equations show that the displacement drives the oscillatory transverse orbital moments, $\langle L^y\rangle$ and $\langle L^z\rangle$, while the last equation shows that these components act as source terms for the uniform $Q^{yz}$-$L^x$ sector. In the time-independent steady state, this source term leaves a finite $\langle L^x\rangle$, and solving the equations to second order reproduces Eqs.~\eqref{mainresult} and \eqref{calAk}.

{\it Numerical Estimation.---} Next, we estimate the magnitude of $\langle L^x_{\bm 0}\rangle$ using $\omega_{\bm q} = v|\bm q|$, $v = 5000 \,{\rm m/s}$, $M = 4 \times 10^{-26}\,{\rm kg}$, and $d = 5 \,\text{\AA}$.
The Debye temperature is approximated as 
$T_D = 240\,{\rm K}$.
We consider the injection of chiral phonons from an adjacent chiral material (see Fig.~\ref{fig:setup}(b)) and assume the nonequilibrium phonon distribution as
$n_{\bm q}^\lambda = n_{{\rm eq},{\bm q}} + \Theta(q_x) \mathcal T / ({e^{\hbar\Omega_{\bm q}^\lambda/k_BT}-1})$ for simplicity, where $k_B$ is the Boltzmann constant, $T$ is the temperature, $\mathcal{T}$ is the transmission coefficient of phonons at the interface, $n_{{\rm eq},{\bm q}}=1/(e^{\hbar \omega_{\bm q}}-1)$, and $\Omega_{\bm q}^\lambda=(\bar{v}' + \lambda \delta v') |\bm q|$ is the phonon dispersion with the chirality $\lambda$ in the adjacent material. In the following estimate, we assume $\bar{v}' = 5000\,{\rm m/s}$, $\delta v' = \bar{v}'/10$, and ${\cal T}=0.1$.
For the sake of numerical convergence, we introduce a broadening factor $\Gamma$ and take it as $0.01\,{\rm eV}$~\footnote{With the broadening factor, $(\epsilon^{x,y}_{\bm k}-\epsilon^z_{\bm k})^{-2}$ is replaced with $\left[(\epsilon^{x,y}_{\bm k}-\epsilon^z_{\bm k})^2 + \Gamma^2\right]^{-1}$.}.

\begin{figure}[tb]
    \centering
    \includegraphics[width=0.8\linewidth]{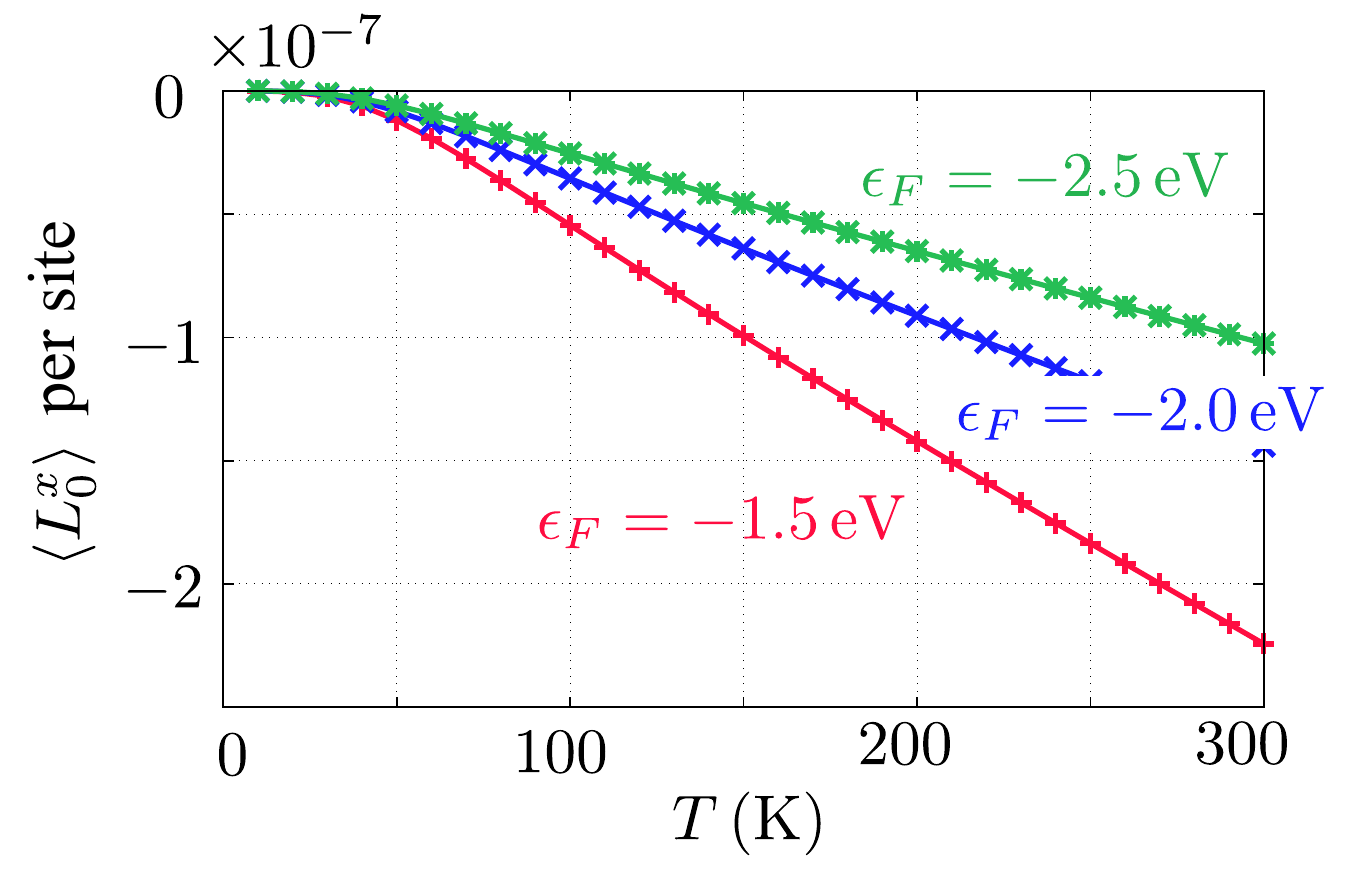}
    \caption{The temperature dependence of $\langle L^x_{\bm 0}\rangle$ is shown.  The unit of the vertical axis is the Bohr magneton per site. 
    $\epsilon_F$ is changed as $-2.5,\,-2.0,\,-1.5\,{\rm eV}$, respectively, for $t_{\sigma} = -1.9 \,{\rm eV}$, and $t_{\pi} = -2.0 \,{\rm eV}$.
    }
    \label{fig:Lx}
\end{figure}

The estimated static orbital accumulation per site, $\langle L^x_{\bm 0}\rangle$, is shown in Fig.~\ref{fig:Lx} as a function of temperature.
As seen in Fig.~\ref{fig:Lx}, $\langle L^{x}_{\bm 0}\rangle$ becomes proportional to the temperature when the temperature is higher than the Debye temperature.
Fig.~\ref{fig:Lx} shows how the orbital accumulation changes when the electron density is varied. As the Fermi energy is raised and the electron density increases, the absolute value of the orbital accumulation is enhanced. This originates from the fact that the orbital accumulation is induced not by the vicinity of the Fermi energy, but by the entire Fermi sea, i.e., all occupied energy eigenstates (see, e.g., Eq.~\eqref{calAk}).

As shown in Fig.~\ref{fig:Lx}, the amplitude of $\langle L_{\bm 0}^x \rangle$ is estimated to be $-8\times 10^{-10} \, [1/{\rm K}] \times T$ for $t_\pi = -2.0 \,{\rm eV}$ and $\epsilon_F = -1.5 \,{\rm eV}$ at high temperatures.
This value of the orbital accumulation is much smaller than the one generated by circularly polarized light, which reaches about $-10\times 10^{-3}$~\cite{adamantopoulos2025light}, because it is proportional to the frequency.
To discuss the efficiency of the generation of the orbital accumulation, we should normalize it with the injected energy flux $I_{\rm phonon}$.
For circular phonon modes, using the high-temperature approximation $I_{\rm phonon} =  (Nd^3)^{-1} \sum_{\bm q,\lambda} \hbar\Omega_{\bm q}^\lambda n_{\bm q}^\lambda (d\Omega_{\bm q}^\lambda / dq_x) \approx \mathcal{T}\bar{v}'k_BT/d^3$~\cite{Waldecker2016}, the efficiency of the orbital accumulation generation is estimated to be $\langle L_{\bm 0}^x\rangle /I_{\rm phonon} \sim -1.5\times 10^{-17}\,{\rm m^2/W}$.
On the other hand, for circularly polarized light, it is estimated $\langle L_{\bm 0}^x\rangle /I_{\rm photon} \sim -1\times 10^{-16}\,{\rm m^2/W}$.
Therefore, the conversion efficiency per unit energy flux is found to be comparable.
Note that the efficiency does not depend on the transmission rate, $\mathcal T$.
The magnitude of the orbital accumulation per site is $\sim 2\times 10^{-7}$ and can be converted into the typical chemical potential difference between orbitals, $\mu_l$, through $\langle L^x_{\bm 0} \rangle_{(2)} \simeq \mu_l D(\epsilon_F)$, where $D(\epsilon)$ is the density of states per site.
In our model, the density of states is approximated as $D(\epsilon_F) \sim 0.6\,{\rm /eV/site}$, and we obtain the orbital chemical potential, $\mu_l \sim 0.4\, \mu{\rm eV}$, which seems to be detectable.

\begin{figure}[tb]
    \centering
\includegraphics[width=0.85\linewidth]{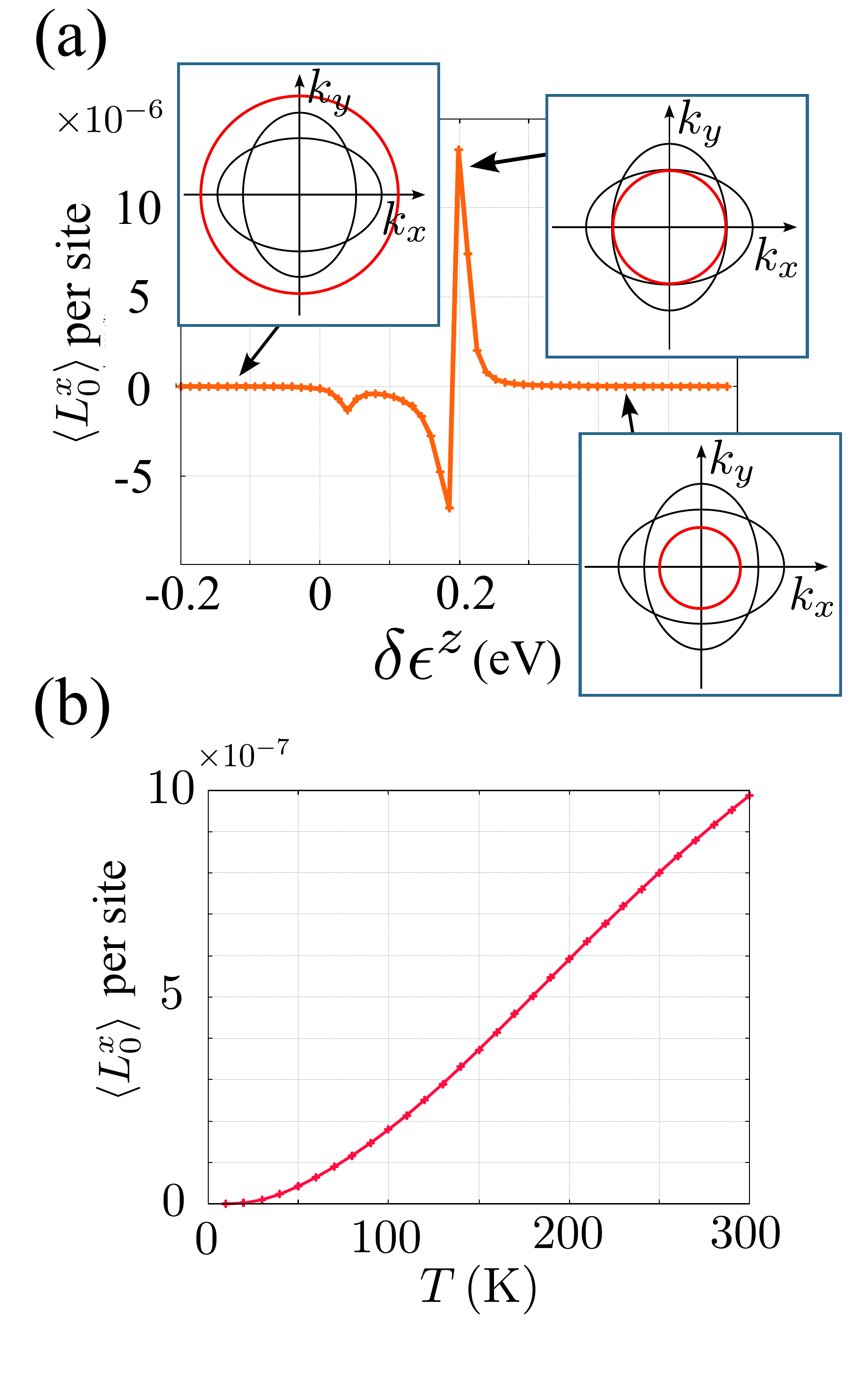}
    \caption{(a) The orbital accumulation, $\langle L^x_0\rangle$, per site is described as a function of $\delta\epsilon^z$.
    The other parameters are taken to be $t_\sigma=-1.9\,{\rm eV}$, $t_\pi = -2.0\,{\rm eV}$, $\epsilon_F=-1.5\,{\rm eV}$, and $T=200\,{\rm K}$.
    The inset shows the schematic of the constant-energy surfaces for each value of $\delta\epsilon_z$.
    Black lines (red line) indicate the energy surface of $p_x$ and $p_y$ orbitals ($p_z$ orbital).
    When $\delta\epsilon_z\sim 0.2\,{\rm eV}$, the $p_z$ band is tangent to the other $p_x$, $p_y$ bands.
    (b) Temperature dependence of the orbital accumulation in $\text{Sr}_2\text{RuO}_4$.}
    \label{fig:degeneracy}
\end{figure}

{\it Effect of Orbital Degeneracy.---} 
As indicated by Eqs.~\eqref{mainresult} and \eqref{calAk}, the orbital accumulation is sensitive to the orbital degeneracy.
To demonstrate the effect of orbital degeneracy, we show the orbital accumulation when shifting the energy of the $p_z$ orbital with $\delta\epsilon_z$ in Fig.~\ref{fig:degeneracy}(a).
The orbital accumulation is sharply enhanced around $\delta\epsilon_z\sim 0.2\,{\rm eV}$, where the Fermi surface of the $p_z$ orbital touches those of the $p_x$ and $p_y$ orbitals, as indicated by the insets.
This result indicates that orbital degeneracies play a key role in generating large orbital accumulation. Such degeneracies naturally arise along high-symmetry directions in the Brillouin zone, giving rise to symmetry-protected orbital moments. For example, in a cubic crystal with $O_h$ symmetry, $d_{xy}$ and $d_{zx}$ orbitals are degenerate along the $\Gamma$-X direction, resulting in the preservation of $L_x$. Consequently, phonon-induced orbital generation is expected to be strongly dependent on crystal direction, similar to electronic orbital injection \cite{Go2023}. Additional band-structure features, such as Dirac or Weyl points, may further enhance orbital generation.

{\it Experimental Relevance.---} To provide a concrete example of orbital accumulation in a realistic material, we consider $\text{Sr}_2\text{RuO}_4$.
This material is effectively described by a square-lattice tight-binding model with $d$ orbitals ($d_{xy}$, $d_{yz}$, and $d_{zx}$), with their dispersions given by $\epsilon_{\bm k}^{yz} = -2t_{2}\cos k_x d - 2t_{1}\cos k_y d - \mu$, $\epsilon_{\bm k}^{zx} = -2t_1 \cos k_x d - 2t_{2}\cos k_y d - \mu$, and $\epsilon_{\bm k}^{xy} = - 2t_3(\cos k_x d + \cos k_y d) - 4t_4 \cos k_x d \cos k_y d - 2t_5(\cos 2k_xd + \cos 2k_y d) - \mu$, where $t_1=88\,{\rm meV}$, $t_2=9\,{\rm meV}$, $t_3=80\,{\rm meV}$, $t_4=40\,{\rm meV}$, $t_5=5\,{\rm meV}$ and $\mu = 109\,{\rm meV}$~\cite{Cobo2016,Philippe2021}. 
The broadening factor, $\Gamma$, is set to $10\,{\rm meV}$.
By neglecting next-nearest-neighbor hopping and assuming $t_1 = t_3$, the electron-phonon interaction is identical in form to Eq.~\eqref{eq:elph} upon the substitutions $t_\sigma \to t_2$ and $t_\pi \to t_1$. 
Using the fact that the orbital angular momentum operator acts on these three d-orbitals as $L_0^x = -i \sum_{\bm k} (c_{\bm k,zx}^\dag c_{\bm k,xy} + {\rm h.c.} )$, the orbital accumulation can be derived by a calculation analogous to Eq.~\eqref{eq:Lxt}.
Fig.~\ref{fig:degeneracy}(b) shows the calculated orbital accumulation as a function of temperature.
The results indicate that the magnitude of the orbital accumulation is comparable to, or even larger than, that obtained in the $p$-orbital model discussed in the main text. 
This demonstrates that our findings are experimentally relevant in realistic systems.

{\it Summary.---} We theoretically formulated and calculated the orbital accumulation generated by circular lattice dynamics, considering surface acoustic waves or nonequilibrium thermal phonons.
We showed that chiral phonons first excite the electron’s orbital quadrupole channel, and that a rectified orbital dipole accumulation emerges at the second order in the lattice displacement. The same static orbital accumulation is obtained consistently within the Berry-phase-based framework and the nonequilibrium Green's-function method.
We also find that the generation of orbital moments by phonons is as efficient as their generation by photons, and we predict that this effect is further enhanced along high-symmetry directions in the Brillouin zone.
Given the broad range of spectral and temporal behavior of phonons, our study opens avenues for the generation of electron-mediated orbital current, both in the adiabatic and ultrafast regimes, in the absence of spin-orbit interaction.

{\it Note added---} Just before submission, we found that a concurrent study (Ref.~\cite{Yao2025b}) discussed orbital accumulation in a honeycomb lattice.

\begin{acknowledgments}
We thank R. Sano and A. Pezo for helpful discussions. This work was supported by Grants-in-Aid for Scientific Research (Grants No. JP23KJ0702 and No. JP24K06951) and the Japan Science and Technology Agency (JST) ASPIRE Program No. JPMJAP2410. A.M. supported by the EIC Pathfinder OPEN grant 101129641 ``OBELIX'', and by France 2030 government investment plan managed by the French National Research Agency under grant reference PEPR SPIN – [SPINTHEORY] ANR-22-EXSP-0009 and [OXIMOR] ANR-24-EXSP-0011.
\end{acknowledgments}

\nocite{SatoData2026}
\bibliography{reference}

\end{document}

\title{Supplementary Information for Orbital Accumulation Induced by Chiral Phonons}

\author{Tetsuya Sato}
\affiliation{Institute for Solid State Physics, University of Tokyo, Kashiwa, 277-8581, Japan}

\author{Takeo Kato}
\affiliation{Institute for Solid State Physics, University of Tokyo, Kashiwa, 277-8581, Japan}

\author{Aurelien Manchon}
\affiliation{Aix-Marseille Univ, CNRS, CINaM, Marseille, France}

\date{\today}

\maketitle

\section{Tight-Binding Model}
\label{app:TBM}

We consider the tight-binding model whose Hamiltonian is given as
\begin{align}
\mathcal{H}_0 &= \sum_{{\bm r},\alpha,\beta} \sum_{{\bm d}} t_{\alpha\beta}({\bm d}) c^\dag_{{\bm r}+{\bm d},\alpha} c_{{\bm r},\beta} , 
\end{align}
where ${c}_{{\bm r},\alpha}$ is an annihilation operator of conduction electrons, $\alpha$ and $\beta$ ($={\rm p}_x,{\rm p}_y,{\rm p}_z$) denote the orbitals, ${\bm e}_x$ and ${\bm e}_y$ are respectively the unit direction vectors pointing to the $x$ and $y$ directions, ${\bm r} = n_x d{\bm e}_x + n_y d{\bm e}_y$ is the lattice vector, $d$ is a lattice constant, and
${\bm d}$ ($= \pm d {\bm e}_x, \pm d{\bm e}_y$) is a displacement vector to the nearest-neighbor sites. 
The orbital-dependent hopping parameter $t_{\alpha\beta}({\bm d})$ is described in the two-center hopping approximation as
\begin{align}
t_{\alpha\beta}({\bm d}) &= \int d^3\bm x \, \phi_\alpha(\bm x-\bm d) H_0 \phi_\beta(\bm x) , \\
H_0 &= -\frac{\hbar^2}{2m}\nabla^2 +V(\bm x-\bm d)+ V(\bm r),
\end{align}
where $\phi_{\alpha}({\bm x})$ is the wavefunction of the atomic orbital $\alpha$.
The Hamiltonian $H_0$ describes the kinetic energy and the Coulomb interaction by ions, where $m$ is the electron mass, $\hbar$ is the reduced Planck's constant, and $V(\bm x)$ is a potential energy produced by a single ion.
For $p$-orbitals, the hopping parameters denoted with $t_{\alpha\beta}({\bm d})$ are reduced by symmetry to two parameters as
\begin{align}
t_\sigma &= \int d^3\bm r \, \phi_x(\bm r - d\bm e_x) \mathcal{H}_0 \phi_x(\bm r)= \int d^3\bm r \, \phi_x(\bm r) H_0' \phi_x(\bm r+ d\bm e_x),\\
t_\pi &= \int d^3\bm r\, \phi_y(\bm r- d\bm e_x) H_0 \phi_y(\bm r )= \int d^3\bm r \, \phi_y(\bm r) H_0' \phi_y(\bm r + d\bm e_x), \\
H_0' &= -\frac{\hbar^2}{2m}\nabla^2 +V(\bm x)+ V(\bm r+\bm d),
\end{align}
where the integral variable is replaced as ${\bm r} \rightarrow {\bm r} + d{\bm e}_x$ in the second equation.
By using the Fourier transformation
\begin{align}
c_{{\bm r},\alpha} &= \frac{1}{\sqrt{N}} \sum_{\bm k} e^{i{\bm k}\cdot {\bm r}}c_{{\bm k},\alpha},\\
c_{{\bm r},\alpha}^\dagger &= \frac{1}{\sqrt{N}} \sum_{\bm k} e^{-i{\bm k}\cdot {\bm r}}c_{{\bm k},\alpha}^\dagger,
\end{align}
with the number of the sites, $N$, the tight-binding model is rewritten in the momentum representation as
\begin{align}
\mathcal{H}_0 &= \sum_{\bm k,\alpha,\beta}  c^\dag_{{\bm k},\alpha} (\hat{h}_{\bm k})_{\alpha,\beta} c_{{\bm k},\beta} ,
\end{align}
where the matrix elements of $\hat{h}_{\bm k}$ are given as
\begin{align}
& (\hat{h}_{\bm k})_{xx} = 2t_{\sigma}\cos k_x d  + 2t_\pi \cos k_y d , \\
& (\hat{h}_{\bm k})_{yy} = 2t_{\pi} \cos k_x d  + 2t_{\sigma}\cos k_y d, \\
& (\hat{h}_{\bm k})_{zz} = 2t_{\pi} (\cos k_x d + \cos k_y d) , \\
& (\hat{h}_{\bm k})_{xy} = (\hat{h}_{\bm k})_{xz} = (\hat{h}_{\bm k})_{yz} = (\hat{h}_{\bm k})_{yx} = (\hat{h}_{\bm k})_{zx} = (\hat{h}_{\bm k})_{zy} = 0.
\end{align}

Next, let us consider the effect of the lattice displacement ${\bm u}({\bm r})$ of the atom located at ${\bm r}$.
Up to the first order of ${\bm u}$, the hopping parameter is modified as
\begin{align}
t_{\alpha\beta}({\bm r},{\bm d}) &\simeq t_{\alpha\beta}({\bm d})
+ {\bm \Lambda}_{\alpha\beta} ({\bm d}) \cdot [{\bm u}({\bm r}+{\bm d}) - {\bm u}({\bm r})] ,\\
{\bm \Lambda}_{\alpha\beta} ({\bm d}) &=\int d^3\bm r \, \phi_\alpha(\bm r) \left[\partial_{\bm r} V(\bm r+\bm d) + H_0' \partial_{\bm r}\right]\phi_\beta(\bm r+\bm d) .
\end{align}
Therefore, the Hamiltonian of the electron-phonon coupling is described as
\begin{align}
{\cal H}_{ep} &= \sum_{{\bm r},\alpha,\beta} \sum_{{\bm d}} {\bm \Lambda}_{\alpha\beta} ({\bm d}) \cdot [{\bm u}({\bm r}+{\bm d})-{\bm u}({\bm r})] c^{\dagger}_{{\bm r}+{\bm d},\alpha}
c_{{\bm r},\beta} .
\end{align}

The diagonal part of the electron-phonon coupling strength, ${\bm \Lambda}_{\alpha\alpha} ({\bm d})$, is described as follows:
\begin{align}
    &{\bm \Lambda}_{xx}(\pm d {\bm e}_x)  = \pm g_1 {\bm e}_x, \\
    &{\bm \Lambda}_{yy}(\pm d {\bm e}_y) = \pm g_1 {\bm e}_y, \\
    &{\bm \Lambda}_{yy}(\pm d {\bm e}_x) = {\bm \Lambda}_{zz}(\pm d {\bm e}_x) = \pm g_2 {\bm e}_x, \\    
    &{\bm \Lambda}_{xx} (\pm d {\bm e}_y) = \bm s_{zz}(\pm d {\bm e}_y) = \pm g_2 {\bm e}_y, 
\end{align}
where $g_1$ and $g_2$ are the integrals defined as
\begin{align}
g_1 &= \int d^3\bm r \phi_x(\bm r)[\partial_x V(\bm r+d\bm e_x)+H_0'\partial_x]\phi_x(\bm r+d\bm e_x), \\
g_2 &= \int d^3\bm r \phi_y(\bm r)[\partial_x V(\bm r+d\bm e_x)+H_0'\partial_x]\phi_y(\bm r+d\bm e_x).
\end{align}
The diagonal part of the electron-phonon coupling corresponds to the ordinary electron-phonon interaction due to the change in bond strengths.
However, it does not contribute to the orbital accumulation in the second-order perturbation with respect to the lattice displacement, since it does not change the orbital degree of freedom.
Hereafter, we will drop this diagonal part.

The off-diagonal part of the electron-phonon coupling strength, ${\bm \Lambda}_{\alpha\beta}({\bm d})$ ($\alpha\ne \beta$), is expressed in terms of the hopping parameters \cite{Mitra1969,Barisic1972,Seungyun2025}, using a rotational operation generated by the orbital angular momentum.
For example, ${\bm \Lambda}_{xy}(d\bm e_x)$ can be rewritten as $\bm e_y (t_\sigma - t_\pi)/d$. To prove this, we rewrite ${\bm \Lambda}_{xy}(d\bm e_x)$ in terms of operators and eigenkets/eigenbras as 
\begin{align}
{\bm \Lambda}_{xy}(d\bm e_x) = \braket{{\rm p}_x | \nabla V(\hat{\bm r}+ d{\bm e}_x)|{\rm p}_y'} +\frac{i}{\hbar} \braket{{\rm p}_x |  \hat{H}_0' \hat{p}_y | {\rm p}_y'} ,
\end{align}
where $\braket{{\bm r}|{\rm p}_\alpha} = \phi_\alpha({\bm r})$ and $\braket{{\bm r}|{\rm p}_\alpha'} = \phi_\alpha({\bm r}+d{\bm e}_x)$  ($\alpha=x,y$) are the wave functions of the atomic orbitals located at the origin and ${\bm r}=-d{\bm e}_x$. The operators of position and momentum are respectively denoted by $\hat{\bm r}$ and $\hat{\bm p}$, and $\hat{H}_0' = \hat{\bm p}^2/2m + V(\hat{\bm r}) + V(\hat{\bm r}+d{\bm e}_x)$.
We define the angular momentum around the origin and ${\bm r} = -d{\bm e}_x$ as $\hat{\bm L} = \hat{\bm r} \times \hat{\bm p}$ and  $\hat{\bm L}' = (\hat{\bm r} + d{\bm e}_x)  \times \hat{\bm p} = \hat{\bm L} + d {\bm e}_x \times  \hat{\bm p}$. Using the spherical symmetry of the Hamiltonian, we obtain
\begin{align}
[\hat{H}_0',\hat{L}_z] &= [{\bm p}^2/2m + V(\hat{\bm r}),\hat{L}_z] + [V(\hat{\bm r}+d{\bm e}_x),\hat{L}_z] \notag \\
&= 0 +  [V(\hat{\bm r}+d{\bm e}_x),\hat{L}_z' - d \hat{p}_y] \notag \\
&= -d [V(\hat{\bm r}+d{\bm e}_x),\hat{p}_y] = -id\hbar \nabla V(\hat{\bm r}+d{\bm e}_x).
\end{align}
Therefore, the $y$-component of ${\bm \Lambda}_{xy} (d{\bm e}_x)$ is calculated as 
\begin{align}
\Lambda_{xy}^y(d\bm e_x) &= \frac{i}{\hbar d} \left[ \braket{{\rm p}_x | [\hat{H}_0',\hat{L}_z]|{\rm p}_y'} +d \braket{{\rm p}_x |  \hat{H}_0' \hat{p}_y | {\rm p}_y'}\right] \notag \\
&=\frac{i}{\hbar d} \left[ 
\braket{{\rm p}_x | \hat{H}_0' (\hat{L}_z + d\hat{p}_y)| {\rm p}_y'} - \braket{{\rm p}_x | \hat{L}_z \hat{H}_0' | {\rm p}_y' } \right] \notag \\
&=\frac{i}{\hbar d} \left[ 
\braket{{\rm p}_x | \hat{H}_0' (\hat{L}_z + d\hat{p}_y)| {\rm p}_y'} - \braket{{\rm p}_x | \hat{L}_z \hat{H}_0' | {\rm p}_y' } \right] \notag \\
&=\frac{i}{\hbar d} \left[ 
\braket{{\rm p}_x | \hat{H}_0' \hat{L}_z'| {\rm p}_y'} - \braket{{\rm p}_x | \hat{L}_z \hat{H}_0' | {\rm p}_y' } \right]. \end{align} 
Finally, using $\hat{L}_z \ket{{\rm p}_x} = i\hbar \ket{{\rm p}_y}$, $\hat{L}_z \ket{{\rm p}_y} = -i\hbar \ket{{\rm p}_x}$, 
$\hat{L}_z' \ket{{\rm p}_x'} = i \hbar \ket{{\rm p}_y'}$,
and $\hat{L}_z' \ket{{\rm p}_y'} = -i\hbar \ket{{\rm p}_x'}$, we obtain 
\begin{align}
\Lambda_{xy}^y(d\bm e_x) &= \frac{1}{d} \left[ \braket{{\rm p}_x | \hat{H}_0' | {\rm p}_x'} - \braket{{\rm p}_y | \hat{H}_0' | {\rm p}_y'}\right]  = \frac{1}{d} (t_\sigma-t_\pi).
\end{align}
We can also show $\Lambda_{xy}^x(d\bm e_x) = \Lambda_{xy}^z(d\bm e_x) = 0$. Thus, the relation ${\bm \Lambda}_{xy}(d\bm e_x)=\bm e_y (t_\sigma - t_\pi)/d$ has been proved.

By the same way, we can also rewrite other components of ${\bm \Lambda}_{\alpha\beta}({\bm d})$.
The results are summarized as
\begin{align}
    &{\bm \Lambda}_{xy} (\pm d\bm e_x) = {\bm \Lambda}_{yx} (\pm d\bm e_x) \simeq \pm \bm e_y (t_\sigma - t_\pi)/d , \label{sxy}
    \\
    &
    {\bm \Lambda}_{xy} (\pm d\bm e_y) = {\bm \Lambda}_{yx} (\pm d\bm e_y) \simeq \pm \bm e_x (t_\sigma - t_\pi)/d,  \label{sxy2}
    \\
    &{\bm \Lambda}_{yz} (\pm d\bm e_y) = {\bm \Lambda}_{zy} (\pm d\bm e_y) \simeq \pm \bm e_z (t_\sigma - t_\pi)/d ,
    \\
    &
    {\bm \Lambda}_{zx} (\pm d\bm e_x) = {\bm \Lambda}_{xz} (\pm d\bm e_x) \simeq \pm \bm e_z (t_\sigma - t_\pi)/d, \\
        &
    {\bm \Lambda}_{yz} (\pm d\bm e_x) = {\bm \Lambda}_{zy} (\pm d\bm e_x) =
    {\bm \Lambda}_{xy} (\pm d\bm e_z) = {\bm \Lambda}_{yx} (\pm d\bm e_z) = 
    {\bm \Lambda}_{zx} (\pm d\bm e_y) = {\bm \Lambda}_{xz} (\pm d\bm e_y) = 0. \label{szero}
\end{align}

Next, we introduce the Fourier transformation
\begin{align}
\bm u({\bm r}) = \frac{1}{N} \sum_{\bm q} e^{i\bm q\cdot\bm r} {\bm u}_{\bm q}.
\end{align}
Since the displacement $\bm u({\bm r})$ is a real vector, $u_{\bm q}^\alpha = (u_{-\bm q}^\alpha)^*$ holds.
Then, the Hamiltonian for the electron-phonon interaction is rewritten as
\begin{align}
& {\cal H}_{ep} = \frac{1}{N} \sum_{{\bm k}, {\bm q},\alpha,\beta} {\bm \Lambda}_{\alpha\beta} ({\bm k},{\bm q}) \cdot {\bm u}_{\bm q} c^{\dagger}_{{\bm k}+{\bm q},\alpha}
c_{{\bm k},\beta} , \\
& {\bm \Lambda}_{\alpha\beta} ({\bm k},{\bm q}) = -\sum_{\bm d} {\bm \Lambda}_{\alpha\beta} ({\bm d}) (e^{-i({\bm k}+{\bm q})\cdot {\bm d}} -e^{-i{\bm q}\cdot {\bm d}}).
\end{align}
Using Eqs.~(\ref{sxy}) and (\ref{sxy2}), the coupling strength ${\bm \Lambda}_{xy}({\bm k},{\bm q})$ is given as
\begin{align}
& {\bm \Lambda}_{xy} ({\bm k},{\bm q}) 
= \frac{2i(t_\sigma-t_\pi)}{d} [ \kappa^x_{{\bm k},{\bm q}} {\bm e}_y
+ \kappa^y_{{\bm k},{\bm q}} {\bm e}_x], \\
& \kappa^\mu_{{\bm k},{\bm q}} = \sin(k_\mu + q_\mu)d - \sin k_\mu d , \quad (\mu = x,y,z).
\end{align}
Other components can be calculated in a similar way.
As a result, the orbital-dependent electron-phonon coupling is derived as
\begin{align}
\mathcal{H}_{ep} = \sum_{\bm k, \bm q} \frac{2i(t_\sigma-t_\pi)}{dN} 
\left\{ \kappa_{\bm k,\bm q}^y u^z_{\bm q} c^\dag_{\bm k+\bm q,y} c_{\bm k,z} + \kappa_{\bm k,\bm q}^x u^z_{\bm q} c^\dag_{\bm k+\bm q,z} c_{\bm k,x} +\left(u^y_{\bm q} \kappa_{\bm k,\bm q}^x + u^x_{\bm q} \kappa_{\bm k,\bm q}^y \right) c^\dag_{\bm k+\bm q,x} c_{\bm k,y} + h.c. \right\}.
\end{align}

Hereafter, we consider the chiral displacement with wavenumber $\pm \bm q$ propagating along $+x$-direction, which is described as 
\begin{align}
\bm u_i(t) = U_{\bm q} \left[\cos(-\omega_{\bm q}t+\bm q\cdot \bm R_i)\bm e_y + \lambda \sin(-\omega_{\bm q}t+\bm q\cdot \bm R_i)\bm e_z\right],
\end{align}
where $U_{\bm q}$ is the amplitude of the displacement, and $\lambda = \pm 1$ denotes the chirality of the phonon.
By the Fourier transformation, this lattice displacement is rewritten as
\begin{align}
    & u_{\bm q}^x=0, \\
    & u_{\bm q}^- \equiv u_{\bm q}^y - i u_{\bm q}^z = \left\{ \begin{array}{ll    
    } 0, & (\lambda = +1), \\ U_{\bm q}e^{-i\omega_{\bm q}t}, & (\lambda = -1), \end{array} \right. \\
    & u_{\bm q}^+ \equiv u_{\bm q}^y + i u_{\bm q}^z = \left\{ \begin{array}{ll} U_{\bm q}e^{-i\omega_{\bm q}t}, & (\lambda = +1 ), \\ 0, & (\lambda = -1).\end{array} \right.
\end{align}
where $(\bm u_{\bm q})^* = \bm u_{-\bm q}$. 
When we assume the long wavelength limit $|\bm k|,|\bm q|\ll 1/d$, and $q_y, q_z \ll q_x$, we obtain
\begin{align}
\label{eq:elph}
\mathcal{H}_{ep} \simeq &
\sum_{\bm k,\bm q} \frac{i\sqrt{2} (t_\sigma-t_\pi) q_x }{N}
\left\{  (u^y_{\bm q} + iu^z_{\bm q}) \left[ - c^\dag_{\bm k+\bm q,x} c_{\bm k,+}  + c^\dag_{\bm k+\bm q,-} c_{\bm k,x} \right] - (u^y_{-\bm q} - iu^z_{-\bm q})  \left[-c^\dag_{\bm k-\bm q,+} c_{\bm k,x} + c^\dag_{\bm k-\bm q,x} c_{\bm k,-} \right] \right\} \nonumber
\\=& \sum_{\bm q}
C_{\bm q} \left[ u_{\bm q}^+  Q_{-\bm q}^-  +  u_{\bm q}^- Q_{-\bm q}^+   \right],
\end{align}
where $C_{\bm q}=-i(t_\sigma-t_\pi)q_x/N$, $c_{\bm k,\pm}=\mp\frac{1}{\sqrt{2}}\left(c_{\bm k,y}\mp i c_{\bm k,z}\right)$,
and $Q_{\bm q}^\pm=\pm\sqrt{2}\sum_{\bm k}( c_{{\bm k},\pm}^\dag  c_{{\bm k}+{\bm q},x} -  c_{\bm k,x}^\dag c_{{\bm k}+{\bm q},\mp})$.
We define quadrupole operators in the real-space representation as 
\begin{align}
& Q_{{\bm r}}^{x^2-y^2} = \sum_{\alpha,\beta} c^\dagger_{{\bm r},\alpha} 
(\hat{L}_x^2 - \hat{L}_y^2)_{\alpha\beta} c_{{\bm r},\beta}, \\
& Q_{{\bm r}}^{3z^2-r^2} = \frac{1}{\sqrt{3}} \sum_{\alpha,\beta}c^\dagger_{{\bm r},\alpha} (3\hat{L}_z^2 - 2) c_{{\bm r},\beta}, \\
& Q_{{\bm r}}^{xy} = \sum_{\alpha,\beta} c^\dagger_{{\bm r},\alpha} (\hat{L}_x \hat{L}_y + \hat{L}_y \hat{L}_x)_{\alpha\beta} c_{{\bm r},\beta}, \\
& Q_{{\bm r}}^{yz} = \sum_{\alpha,\beta} c^\dagger_{{\bm r},\alpha} (\hat{L}_y \hat{L}_z + \hat{L}_z \hat{L}_y)_{\alpha\beta} c_{{\bm r},\beta}, \\
& Q_{{\bm r}}^{zx} = \sum_{\alpha,\beta} c^\dagger_{{\bm r},\alpha} 
(\hat{L}_z  \hat{L}_x+\hat{L}_x \hat{L}_z)_{\alpha\beta} c_{{\bm r},\beta}, \\
&\hat{L}_x = \hbar \left( \begin{array}{ccc} 0 & 0 & 0 \\ 0 & 0 & -i \\ 0 & i & 0 \end{array} \right),  \\
&\hat{L}_y = \hbar \left( \begin{array}{ccc} 0 & 0 & i \\ 0 & 0 & 0 \\ -i & 0  & 0 \end{array} \right),  \\
&\hat{L}_z = \hbar \left( \begin{array}{ccc} 0 & -i & 0 \\ i & 0 & 0 \\ 0 & 0 & 0 \end{array} \right) .
\end{align}
The Fourier transformation of the quadrupole operators is easily expressed using electron annihilation and creation operators.
For example, $Q_{\bm q}^{xy}$ is calculated as 
\begin{align}
Q_{\bm q}^{xy} = \sum_{\alpha,\beta} c^\dagger_{{\bm k},\alpha} (\hat{L}_x \hat{L}_y + \hat{L}_y \hat{L}_x)_{\alpha\beta} c_{{\bm k}+{\bm q},\beta}.
\end{align}
The other quadrupole operators are described in a similar way.
$Q_{\bm q}^\pm$ can be expressed as $Q_{\bm q}^\pm = Q_{\bm q}^{xy}\pm i Q_{\bm q}^{zx}$.

\section{Detailed Calculation of the Berry curvature}

In this appendix, a detailed calculation of the orbital angular momentum based on the Berry curvature is presented, assuming classical lattice displacement.
If the lattice displacement is introduced adiabatically as a classical external field, the Hamiltonian of the whole system can be written as
\begin{align}
    &\mathcal{H} = \mathcal{H}_{\rm el} + \mathcal{H}_{\rm ep} + \mathcal{H}_Z,
    \\
    &\mathcal{H}_{\rm el} = \sum_{\bm k ,\alpha=(x,y,z)} \epsilon_{\bm k}^\alpha c_{\bm k\alpha}^\dag c_{\bm k\alpha},\,\,
    \\
    &\mathcal{H}_{\rm ep}= \sum_{\bm q} C_{\bm q} [u_{\bm q}^+ Q_{-\bm q}^- + u_{\bm q}^- Q_{-\bm q}^+],\,\,
    \\
    &\mathcal{H}_Z=\gamma \sum_{\bm q} \bm B_{-\bm q}^\lambda \cdot \bm L_{\bm q}^\lambda,
\end{align}
where $\mathcal{H}_Z$ describes an auxiliary Zeeman field.
The expectation value of the orbital angular momentum can be written using $\mathcal{H}_Z$ as
\begin{align}
    \langle L^\mu_{\bm q} \rangle(t) =  \frac{1}{\gamma} \left.{\rm Tr}\left\{ \rho(t) \left(\partial_{B_{-\bm q}^\mu} \mathcal{H}\right)\right\}\right|_{\bm B_{\bm q}\rightarrow 0},
\label{eq:L_berry}
\end{align}
where $\rho(t)$ is the electronic density matrix.
The time evolution of the density matrix follows the von Neumann equation given as
\begin{align}
    i\hbar\partial_t \rho(t) = [\mathcal{H},\rho(t)]  .
\label{eq:Neumann}
\end{align}
When the lattice displacement is sufficiently slow, the diagonal part of the density matrix can be written in terms of the eigenstates $\ket{\psi_\alpha}$ of $\mathcal{H}$ for given $\bm B$ and $\bm u$ as
\begin{align}
    \rho_{\rm diag}(t) = \sum_\alpha f^\alpha \ket{\psi_\alpha}\bra{\psi_\alpha}.
\label{eq:rho_eq}
\end{align}
Note that the eigenstates $\ket{\psi_\alpha}$ depend on the lattice displacement and therefore differ from the usual unperturbed states $\ket{\bm k \alpha}$ ($\alpha=x,y,z$) determined by $\mathcal{H}_{\rm el}$.
Writing the off-diagonal part of the density matrix as $\delta\rho(t)$, the density matrix is expanded as
\begin{align}
    \rho(t) = \rho_{\rm diag}(t) + \delta\rho(t).
\end{align}
From the von Neumann equation \eqref{eq:Neumann} for $\alpha \neq \beta$, we obtain
\begin{align}
    \bra{\psi_{\alpha}}\delta\rho(t)\ket{\psi_{\beta}} &= \frac{\bra{\psi_{\alpha}}\mathcal{H}\delta\rho(t)-\delta\rho(t)\mathcal{H}\ket{\psi_{\beta}}}{\epsilon^\alpha-\epsilon^\beta} = \frac{\bra{\psi_{\alpha}}\mathcal{H}\rho(t)-\rho(t)\mathcal{H}\ket{\psi_{\beta}}}{\epsilon^{\alpha}-\epsilon^{\beta}} \nonumber
    \\&= \frac{\bra{\psi_{\alpha}}i\hbar\partial_t\rho(t)\ket{\psi_{\beta}}}{\epsilon^{\beta}-\epsilon^{\beta}} 
    = \sum_{\bm q}\dot{u}_{\bm q}^\lambda \frac{\bra{\psi_{\alpha}}i\hbar\partial_{u_{\bm q}^\lambda}\rho(t)\ket{\psi_{\beta}}}{\epsilon^{\alpha}-\epsilon^{\beta}} 
    \simeq \sum_{\bm q}\dot{u}_{\bm q}^\lambda  \frac{\bra{\psi_{\alpha}}i\hbar\partial_{u_{\bm q}^\lambda}\rho_{eq}(t)\ket{\psi_{\beta}}}{\epsilon^{\alpha}-\epsilon^{\beta}} ,
\end{align}
where $\epsilon^{\alpha}$ is the eigenvalue satisfying $\mathcal{H}\ket{\psi_{\alpha}} = \epsilon^{\alpha}\ket{\psi_{\alpha}}$.
Substituting Eq.~\eqref{eq:rho_eq} yields
\begin{align}
    \bra{\psi_{\alpha}}\delta\rho(t)\ket{\psi_{\beta}}&= \sum_{\bm q} i\hbar\dot{u}_{\bm q}^\lambda \frac{f^{\beta} \braket{\psi_{\alpha}|\partial_{u_{\bm q}^\lambda}\psi_{\beta}} + f^{\alpha} \braket{\partial_{u_{\bm q}^\lambda}\psi_{\alpha}|\psi_{\beta}}}{\epsilon^{\alpha}-\epsilon^{\beta}},
\end{align}
Inserting this result into the definition of the orbital angular momentum \eqref{eq:L_berry}, we obtain
\begin{align}
\label{eq:berry_L}
    \langle L_{\bm q}^\mu\rangle(t) 
    &=  
    \frac{1}{\gamma} \left.{\rm Tr}\left\{ \rho(t) \left(\partial_{B_{-\bm q}^\mu} \mathcal{H}\right)\right\}\right|_{\bm B_{\bm q}\rightarrow 0}
    = \langle L_{\bm q}^\mu\rangle_{\rm diag} + \langle L_{\bm q}^\mu\rangle_{\rm geo} , \\
    \langle L_{\bm q}^\mu \rangle_{\rm diag} &= \frac{1}{\gamma} {\rm Tr}\left\{ \rho_{\rm diag}(t) L_{\bm q}^\mu\right\} = \sum_{\alpha} f^{\alpha} \bra{\psi_{\alpha}}L_{\bm q}^\mu\ket{\psi_{\alpha}},\\
    \langle L_{\bm q}^\mu\rangle_{\rm geo} &= \frac{1}{\gamma} {\rm Tr}\left\{ \delta\rho(t) L_{\bm q}^\mu\right\} ,
\end{align}
The geometrical term, in particular, is described by the Berry curvature, as discussed later.

We calculate the expectation value of the orbital moment up to the first order.
The orbital moment originating from the diagonal part vanishes as
\begin{align}
    \sum_{\alpha} f_\alpha \bra{\psi_\alpha} L_{\bm q}^\pm \ket{\psi_\alpha} &= \sum_{\bm k,\alpha,\beta} C_{\bm q} f_\alpha \frac{\bra{\bm k \alpha} L^\pm_{\bm q} \ket{\bm k' \beta}\! \bra{\bm k' \beta} Q^{\bar\lambda}_{-{\bm q}} \ket{\bm k \alpha} + \bra{\bm k \alpha} Q^{\bar\lambda}_{-{\bm q}}\ket{\bm k' \beta}\!\bra{\bm k' \beta} L^\pm_{\bm q} \ket{\bm k \alpha}}{\epsilon_{\bm k}^\alpha - \epsilon_{\bm k'}^\beta}
    = 0 ,
\end{align}
where $Q_{{\bm q}}^- = (Q_{{\bm q}}^+)^\dag=-\sqrt{2}\sum_{\bm k}(c_{\bm k,-}^\dag c_{\bm k +\bm q,x}- c_{\bm k,x}^\dag c_{\bm k +\bm q,+})$ and $L_{\bm q}^\pm = \sqrt{2}\sum_{\bm k}(c_{\bm k,\pm}^\dag c_{\bm k + \bm q,x} + c_{\bm k,x}^\dag c_{\bm k + \bm q,\mp})$.
We have also used the perturbed state given as
\begin{align}
\nonumber
\ket{\psi_\alpha} = &\ket{\bm k \alpha}+ \sum_{\bm k' \beta} \frac{\bra{\bm k' \beta}\mathcal{H}_{\rm ep}\ket{\bm k \alpha}}{\epsilon_{\bm k}^\alpha-\epsilon_{\bm k'}^\beta} \ket{\bm k' \beta} 
\\\nonumber
&-\frac{1}{2}\sum_{\bm k' \beta} \frac{\bra{\bm k \alpha}\mathcal{H}_{\rm ep}\ket{\bm k' \beta} \! \bra{\bm k' \beta}\mathcal{H}_{\rm ep}\ket{\bm k \alpha}}{(\epsilon_{\bm k}^\alpha - \epsilon_{\bm k'}^\beta)^2} \ket{\bm k \alpha} 
\\
&+ \sum_{\bm k' \beta} \Bigg(\sum_{\bm k''\gamma} \frac{\bra{\bm k'\beta}\mathcal{H}_{\rm ep}\ket{\bm k'' \gamma}\! \bra{\bm k'' \gamma}\mathcal{H}_{\rm ep}\ket{\bm k \alpha}}{(\epsilon_{\bm k}^\alpha-\epsilon_{\bm k''}^\gamma)(\epsilon_{\bm k}^\alpha-\epsilon_{\bm k'}^\beta)} 
- \frac{\bra{\bm k \alpha}\mathcal{H}_{\rm ep}\ket{\bm k \alpha} \! \bra{\bm k'\beta}\mathcal{H}_{\rm ep}\ket{\bm k \alpha}}{(\epsilon_{\bm k}^\alpha - \epsilon_{\bm k'}^\beta)^2}\Bigg) \ket{\bm k' \beta}. 
\end{align}
We note that the terms of $\bra{\bm k \alpha} L_{\bm q}^+\ket{\bm k'\beta}\! \bra{\bm k'\beta} Q_{\bm q}^+\ket{\bm k \alpha}$ vanish since they do not preserve the wavenumber.
Thus, the orbital moment is fully determined from the geometrical part, and it is calculated as
\begin{align}
\nonumber
    \langle L_{\bm q}^\pm \rangle_{(1)}(t) 
    &= - i\hbar \sum_{\bm k, \alpha,\beta} (f_\alpha - f_\beta) C_{\bm q} \dot u_{\bm q}^\lambda \frac{\bra{\bm k \alpha} L_{\bm q}^\pm \ket{\bm k' \beta} \! \bra{\bm k'\beta} Q_{-{\bm q}}^{\bar\lambda} \ket{\bm k \alpha}}{(\epsilon_{\bm k}^\alpha-\epsilon_{\bm k'}^\beta)^2}
    \\\nonumber&\simeq - i\hbar C_{\bm q} \dot u_{\bm q}^\lambda \sum_{\bm k} \Bigg\{ (f_x - f_y) \frac{\bra{\bm k x} L_{\bm q}^\pm \ket{\bm k' y} \! \bra{\bm k' y} Q_{-{\bm q}}^{\bar\lambda} \ket{\bm k x}}{(\epsilon_{\bm k}^x-\epsilon_{\bm k}^y)^2} 
    +(f_x - f_z) \frac{\bra{\bm k x} L_{\bm q}^\pm \ket{\bm k' z} \! \bra{\bm k' z} Q_{-{\bm q}}^{\bar\lambda} \ket{{\bm k} x}}{(\epsilon_{\bm k}^x-\epsilon_{\bm k}^z)^2}
    \\\nonumber&\hspace{2cm}+(f_y - f_x) \frac{\bra{\bm k y} L_{\bm q}^\pm \ket{\bm k' x} \! \bra{\bm k' x} Q_{-{\bm q}}^{\bar\lambda} \ket{\bm k y}}{(\epsilon_{\bm k}^y-\epsilon_{\bm k}^x)^2}
    +(f_z - f_x) \frac{\bra{\bm k z} L_{\bm q}^\pm \ket{\bm k' x} \! \bra{\bm k' z} Q_{-{\bm q}}^{\bar\lambda} \ket{\bm k z}}{(\epsilon_{\bm k}^z-\epsilon_{\bm k}^x)^2}
    \Bigg\}
    \\\nonumber&= -\hbar\omega_{\bm q} C_{\bm q} u_{\bm q}^\lambda \sum_{\bm k} \Bigg\{ \mp \frac{f_x - f_y}{(\epsilon_{\bm k}^x-\epsilon_{\bm k}^y)^2} - \lambda \frac{f_x - f_z}{(\epsilon_{\bm k}^x-\epsilon_{\bm k}^z)^2}
    \pm \frac{f_y - f_x}{(\epsilon_{\bm k}^y-\epsilon_{\bm k}^x)^2} + \lambda \frac{f_{\bm k}^z - f_{\bm k}^x}{(\epsilon_{\bm k}^z-\epsilon_{\bm k}^x)^2} \Bigg\}
    \\&
    = 2\hbar\omega_{\bm q} C_{\bm q} u_{\bm q}^\lambda \sum_{\bm k} \Bigg\{ \lambda \frac{f_{\bm k}^x-f_{\bm k}^z}{(\epsilon_{\bm k}^x-\epsilon_{\bm k}^z)^2} \pm \frac{f_{\bm k}^x-f_{\bm k}^y}{(\epsilon_{\bm k}^x-\epsilon_{\bm k}^y)^2} \Bigg\}.
\label{app:eq:Lq}
\end{align}
In addition, we note that $\langle L^x_{\bm q} \rangle$ is zero up to the first order because we neglected the mixing interaction between $p_y$ and $p_z$ orbitals based on the assumption of $|q_y|\ll |q_x|$.
These results are consistent with the findings of linear response theory, as shown later.

In the same manner, we obtain the orbital accumulation, $\langle L^x_{\bm 0}\rangle$, in the second order with respect to the displacement.
Regarding the diagonal part, they vanish by assuming $|\bm q|\ll|\bm k|$ (See Eq.~\eqref{eq:Keldysh_result} for details).
The geometrical part is calculated as
\begin{align}
\nonumber
    \langle L^x_{\bm 0} \rangle_{(2)} 
    &\simeq \sum_{\bm k,\bm k'} \hbar\omega_{\bm q} u_{\bm q}^\lambda u_{-\bm q}^{\bar\lambda}  |C_{\bm q}|^2 \Bigg\{ \Bigg[ \sum_{\alpha,\beta} f_{\bm k}^x \frac{\bra{\bm kx} Q_{-\bm q}^\lambda \ket{\bm k'\beta} \bra{\bm k'\beta} L^x_0 \ket{\bm k'\alpha} }{\epsilon_{\bm k}^x-\epsilon_{\bm k}^\beta} \frac{ \bra{\bm k'\alpha} Q_{\bm q}^{\bar\lambda}  \ket{\bm kx}}{(\epsilon_{\bm k}^\alpha-\epsilon_{\bm k}^x)(\epsilon_{\bm k}^x-\epsilon_{\bm k}^\alpha)} 
    \\\nonumber&\hspace{5mm}+ f_{\bm k}^y \frac{\bra{\bm ky} L^x_0 \ket{\bm kz}\bra{\bm kz} Q_{-\bm q}^\lambda \ket{\bm k'x}}{\epsilon_{\bm k}^x-\epsilon_{\bm k}^z} \frac{ \bra{\bm k'x} Q_{\bm q}^{\bar\lambda} \ket{\bm ky}}{(\epsilon_{\bm k}^x-\epsilon_{\bm k}^y)(\epsilon_{\bm k}^y-\epsilon_{\bm k}^x)}
    + f_{\bm k}^y \bra{\bm ky} L^x_0 \ket{\bm kz} \frac{\bra{\bm kz} Q_{-\bm q}^\lambda \ket{\bm k'x}}{\epsilon_{\bm k}^z-\epsilon_{\bm k}^x} \frac{ \bra{\bm k'x} Q_{\bm q}^{\bar\lambda} \ket{\bm ky}}{(\epsilon_{\bm k}^z-\epsilon_{\bm k}^y)(\epsilon_{\bm k}^y-\epsilon_{\bm k}^x)}
    \\\nonumber&\hspace{5mm}+ f_{\bm k}^y \bra{\bm ky} L^x_0 \ket{\bm kz} \frac{ \bra{\bm kz} Q_{-\bm q}^\lambda \ket{\bm k'x} \bra{\bm k'x} Q_{\bm q}^{\bar\lambda} \ket{\bm ky}}{(\epsilon_{\bm k}^z-\epsilon_{\bm k}^y)(\epsilon_{\bm k}^y-\epsilon_{\bm k}^z)(\epsilon_{\bm k}^y-\epsilon_{\bm k}^x)}
    + f_{\bm k}^y \bra{\bm ky} L^x_0 \ket{\bm kz} \frac{ \bra{\bm kz} Q_{\bm q}^{\bar\lambda} \ket{\bm k'x} \bra{\bm k'x} Q_{-\bm q}^\lambda \ket{\bm ky}}{(\epsilon_{\bm k}^z-\epsilon_{\bm k}^y)(\epsilon_{\bm k}^y-\epsilon_{\bm k}^z)(\epsilon_{\bm k}^y-\epsilon_{\bm k}^x)}
    \\\nonumber&\hspace{5mm}+ f_{\bm k}^z \frac{\bra{\bm kz} L^x_0 \ket{\bm ky}\bra{\bm ky} Q_{-\bm q}^\lambda \ket{\bm k'x}}{\epsilon_{\bm k}^x-\epsilon_{\bm k}^y} \frac{ \bra{\bm k'x} Q_{\bm q}^{\bar\lambda} \ket{\bm kz}}{(\epsilon_{\bm k}^x-\epsilon_{\bm k}^z)(\epsilon_{\bm k}^z-\epsilon_{\bm k}^x)}
    + f_{\bm k}^z \bra{\bm kz} L^x_0 \ket{\bm ky} \frac{\bra{\bm ky} Q_{-\bm q}^\lambda\ket{\bm k'x}}{\epsilon_{\bm k}^y-\epsilon_{\bm k}^x} \frac{ \bra{\bm k'x} Q_{\bm q}^{\bar\lambda} \ket{\bm kz}}{(\epsilon_{\bm k}^y-\epsilon_{\bm k}^z)(\epsilon_{\bm k}^z-\epsilon_{\bm k}^x)}
    \\\nonumber&\hspace{5mm}+ f_{\bm k}^z \bra{\bm kz} L^x_0 \ket{\bm ky} \frac{ \bra{\bm ky} Q_{-\bm q}^\lambda \ket{\bm k'x} \bra{\bm k'x} Q_{\bm q}^{\bar\lambda} \ket{\bm kz}}{(\epsilon_{\bm k}^y-\epsilon_{\bm k}^z)(\epsilon_{\bm k}^z-\epsilon_{\bm k}^y)(\epsilon_{\bm k}^z-\epsilon_{\bm k}^x)}
    + f_{\bm k}^z \bra{\bm kz}  L^x_0 \ket{\bm ky} \frac{ \bra{\bm ky} Q_{\bm q}^{\bar\lambda} \ket{\bm k'x} \bra{\bm k'x} Q_{-\bm q}^\lambda \ket{\bm kz}}{(\epsilon_{\bm k}^y-\epsilon_{\bm k}^z)(\epsilon_{\bm k}^z-\epsilon_{\bm k}^y)(\epsilon_{\bm k}^z-\epsilon_{\bm k}^x)} \Bigg] 
    \\\nonumber&\hspace{5mm}- (\lambda \leftrightarrow {\bar\lambda} ) + h.c. \Bigg\} 
    \\\nonumber& = \sum_{\bm k} 4\lambda U_{\bm q}^2 \hbar\omega_{\bm q}  |C_{\bm q}|^2 \Bigg\{  f_{\bm k}^x \frac{1}{(\epsilon_{\bm k}^x-\epsilon_{\bm k}^y)(\epsilon_{\bm k}^x-\epsilon_{\bm k}^z)^2} + f_{\bm k}^x \frac{1}{(\epsilon_{\bm k}^x-\epsilon_{\bm k}^z)(\epsilon_{\bm k}^x-\epsilon_{\bm k}^y)^2} 
    \\
    &\hspace{4cm}- f_{\bm k}^y \frac{1}{(\epsilon_{\bm k}^y-\epsilon_{\bm k}^z)(\epsilon_{\bm k}^y-\epsilon_{\bm k}^x)^2} - f_{\bm k}^z \frac{1}{(\epsilon_{\bm k}^z-\epsilon_{\bm k}^y)(\epsilon_{\bm k}^z-\epsilon_{\bm k}^x)^2} \Bigg\} . \label{app:eq:Lq2nd}
\end{align}

The same result is obtained by the Berry-phase-based formulation~\cite{Yao2025}, in which the orbital angular momentum is expressed as 
\begin{align}
\label{eq:Berry_general}
\langle \bm L_{\bm q}\rangle = \sum_\alpha f_\alpha \bra{\psi_\alpha} \bm L_{\bm q} \ket{\psi_\alpha} + \frac{\hbar}{\gamma} \sum_{\alpha} f_\alpha  \dot{u}_{\bm q}^\lambda \tilde B_{\bm B_{-\bm q},u^{\lambda}_{\bm q}} \Bigg|_{\bm B= 0},
\end{align}
with the assumption that $|\bm q|\ll|\bm k|$, where $\tilde B_{\eta,\eta'}$ ($\eta,\eta' = {\bm B}_{\bm q},u_{\bm q}^\lambda$) is the Berry curvature defined as $\tilde A_\eta = -i\bra{\psi_\alpha}\ket{\partial_\eta \psi_\alpha},\, \tilde B_{\eta,\eta'} = \partial_\eta \tilde A_{\eta'} - \partial_{\eta'} \tilde A_{\eta}$. 
Then, the orbital angular momentum can be derived systematically order by order in the lattice displacement, which enables us to treat the first- and second-order responses in a unified manner.
The linear response of the orbital moment, $\langle \bm L_{\bm q}\rangle_{(1)}$, against the coherent displacement is calculated by taking $\bm u=0$ in Eq.~\eqref{eq:Berry_general}.
Using the Hamiltonian \eqref{eq:elph}, we obtain the same expression as Eq.~\eqref{app:eq:Lq} for the linear response of the orbital moment, with $\langle L^x_{\bm q} \rangle_{(1)} = 0$~\footnote{$\langle L^x_{\bm q}\rangle$ is not prohibited from the symmetry but vanishes because we neglected the mixing interaction between $p_y$ and $p_z$ orbitals due to the assumption of $q_y\ll q_x$.}.
This orbital moment is proportional to the displacement $u_{\bm q}^\lambda(t)$ and oscillates temporally with the phonon frequency $\omega_{\bm q}$, leading to a vanishing orbital moment upon temporal average.
Using the Berry-phase-based formula~Eq.~\eqref{eq:Berry_general}, we can also calculate the second-order respons and obtain the same expression given in Eq.~\eqref{app:eq:Lq2nd}.
This result iscalculation method and  
which reproduces the results of the main article and is also equivalent to Eqs.~(23) and (B5) of Ref.~\cite{Yao2025}, except for the distribution function of the electrons.

\section{Linear Response Theory}
\label{app:LRT}

Here, we consider linear response theory to derive the orbital moment while assuming the classical displacement modes.
The linear response theory leads to
\begin{align}
\nonumber
&\langle L^\pm_{\bm q} \rangle (t) =  \langle L^\pm_{\bm q} \rangle^0 - \frac{i}{\hbar} C_{\bm q} \int_{-\infty}^t dt' \Bigg\{  \langle [ L^\pm_{\bm q}(t),  Q_{-\bm q}^-(t')] \rangle  u_{\bm q}^+(t') 
+ \langle [ L^\pm_{\bm q}(t),  Q_{-\bm q}^+(t')] \rangle u_{\bm q}^-(t') \Bigg\},
\end{align}
using the wave number conservation.
We note that $\langle L^x_{\bm q}\rangle $ vanishes since the hopping between $p_y$ and $p_z$ orbitals is neglected in our electron-phonon Hamiltonian.
The retarded response function is defined as 
\begin{align}
    \phi_{\bm q}^{\pm,\lambda} (t-t') = \frac{i}{\hbar} \theta(t-t') \langle [L_{\bm q}^\pm(t-t'),Q_{-\bm q}^\lambda(0)]\rangle.
\end{align}
When we consider the chiral displacement, $u_{\bm q}^\lambda = U_{\bm q} e^{i\omega_{\bm q}t}$ with $u_{\bm q}^{\bar\lambda}=0$, where $\bar\lambda$ is the opposite sign of $\lambda$, the orbital moment, $\langle L^\pm_{\bm q} \rangle$, becomes
\begin{align}
    &\langle L^\pm_{\bm q} \rangle (t) = - C_{\bm q} \int_{-\infty}^\infty dt' \phi_{\bm q}^{\pm,\bar\lambda} (t-t') u_{\bm q}^\lambda(t').
\end{align}
In the frequency space, it reads
\begin{align}
&    \langle L^\pm_{\bm q} \rangle (\omega) = - C_{\bm q}\chi^{\pm,\bar\lambda}_{\bm q}(\omega)u^\lambda_{\bm q}(\omega),
\end{align}
where
\begin{align}
    u^\lambda_{\bm q}(t) &= \int_{-\infty}^\infty \frac{d\omega}{2\pi} u^\lambda_{\bm q}(\omega) e^{-i\omega t+\delta t},\\   
\chi^{\pm,\bar\lambda}_{\bm q}(\omega) &= \int_{-\infty}^t \phi_{\bm q}^{\pm,\bar\lambda} (t-t') e^{i\omega(t-t')-\delta(t-t')} dt'.
\end{align}
The corresponding response function of imaginary time reads
\begin{align}
    \Phi_{\bm q}^{\pm,\bar\lambda} (\tau-\tau') = \langle T_\tau L_{\bm q}^\pm(\tau) Q_{-\bm q}^{\bar\lambda}(\tau') \rangle,
\end{align}
and it can be calculated using the Feynman diagram method as
\begin{align}
\nonumber
    \Phi_{\bm q}^{\pm \bar\lambda} (\tau-\tau') 
    &= 2 \sum_{\bm k,\bm k'} \Bigg\langle T_\tau (c_{\bm k-\bm q,\pm}^\dag(\tau) c_{\bm k,x}(\tau) + c_{\bm k-\bm q,x}^\dag(\tau) c_{\bm k,\mp}(\tau)) 
     (-\lambda c_{\bm k',\bar\lambda}^\dag(\tau') c_{\bm k'-\bm q,x}(\tau') + \lambda c_{\bm k',x}^\dag(\tau') c_{\bm k'-\bm q,\lambda}(\tau')) \Bigg\rangle
    \\&
    = 2 \lambda \sum_{\bm k} \Bigg\{ \mathcal{G}^{\mp\bar\lambda}_{\bm k}(\tau-\tau') \mathcal{G}^{xx}_{\bm k-\bm q}(\tau'-\tau) 
    - \mathcal{G}^{+\lambda}_{\bm k-\bm q}(\tau'-\tau) \mathcal{G}^{xx}_{\bm k}(\tau-\tau') \Bigg\},
\end{align}
where the former factor $2$ comes from the definition of $L$.
The Fourier transformation leads to
\begin{align}
\nonumber
    \Phi_{\bm q}^{\pm\bar\lambda} (i\omega_n) &= \int_0^\beta d\tau e^{i\omega_n\tau} \Phi_{\bm q}^{\pm \bar\lambda} (\tau) 
    \\
    &= -2 \lambda k_B T \sum_{\bm k,m} \Bigg\{- \mathcal{G}^{\mp \bar\lambda}_{\bm k}(i\epsilon_m) \mathcal{G}^{xx}_{\bm k-\bm q}(i\epsilon_m-i\omega_n) 
    + \mathcal{G}^{+ \lambda}_{\bm k-\bm q}(i\epsilon_m-i\omega_n) \mathcal{G}^{xx}_{\bm k}(i\epsilon_m) \Bigg\},
\end{align}
where $\omega_m = 2m\pi k_BT$ is the Matsubara frequency for the Bose particle.
By summing over the Matsubara frequency, it becomes
\begin{align}
\nonumber
    &\Phi_{\bm q}^{\pm \bar\lambda} (i\omega_n) 
    = 2 \lambda \sum_{\bm k} \int_C \frac{dz}{2\pi i} F(z) \Bigg\{ -\mathcal{G}^{\mp \bar\lambda}_{\bm k}(z) \mathcal{G}^{xx}_{\bm k-\bm q}(z-i\omega_n) 
    + \mathcal{G}^{+ \lambda}_{\bm k-\bm q}(z-i\omega_n) \mathcal{G}^{xx}_{\bm k}(z) \Bigg\},
\end{align}
where $F(z) = 1/(e^{\beta z}+1)$.
Since $\pm$ is not the eigenstate of the electrons, we rewrite the Green's function in terms of $x,y,z$ as
\begin{align}
    \mathcal{G}_{\bm k}^{xx}(z) &= \frac{1}{z-\epsilon_{\bm k}^x},\\
    \mathcal{G}_{\bm k}^{-+}(z) &= -\frac{1}{2}(\mathcal{G}_{\bm k}^{y}(z) - \mathcal{G}_{\bm k}^{z}(z) ), \\
    \mathcal{G}_{\bm k}^{--}(z) &= \frac{1}{2} (\mathcal{G}_{\bm k}^{y}(z) + \mathcal{G}_{\bm k}^{z}(z)), 
    \\
    \mathcal{G}_{\bm k}^{+-}(z) &= - \frac{1}{2}(\mathcal{G}_{\bm k}^{y}(z) - \mathcal{G}_{\bm k}^{z}(z) ), \\
    \mathcal{G}_{\bm k}^{++}(z) &= \frac{1}{2} (\mathcal{G}_{\bm k}^{y}(z) + \mathcal{G}_{\bm k}^{z}(z)) ,
\end{align}
and by substituting the bare Green's function, we obtain 
\begin{align}
\nonumber
    \Phi_{\bm q}^{\pm \bar\lambda} (i\omega_n) 
    \nonumber
    &=  \sum_{\bm k} \int_C \frac{dz}{2\pi i} F(z) 
    \\\nonumber
    &\hspace{5mm} \times\Bigg\{ -\left(\pm\frac{1}{z-\epsilon_{\bm k}^y} + \lambda \frac{1}{z-\epsilon_{\bm k}^z} \right) \frac{1}{z-i\omega_n-\epsilon_{\bm k-\bm q}^x} 
    + \left(\pm\frac{1}{z-i\omega_n-\epsilon_{\bm k-\bm q}^y} + \lambda \frac{1}{z-i\omega_n-\epsilon_{\bm k-\bm q}^z} \right)  \frac{1}{z-\epsilon_{\bm k}^x} \Bigg\}
    \\
    &= - \sum_{\bm k} \Bigg\{ - \lambda\frac{f(\epsilon_{\bm k-\bm q}^x)-f(\epsilon_{\bm k}^z)}{i\omega_n+\epsilon_{\bm k-\bm q}^x-\epsilon_{\bm k}^z} \mp \frac{f(\epsilon_{\bm k-\bm q}^x)-f(\epsilon_{\bm k}^y)}{i\omega_n+\epsilon_{\bm k-\bm q}^x-\epsilon_{\bm k}^y}  
    -  \lambda \frac{f(\epsilon_{\bm k}^x) - f(\epsilon_{\bm k-\bm q}^z)}{i\omega_n+\epsilon_{\bm k-\bm q}^z-\epsilon_{\bm k}^x} \mp \frac{f(\epsilon_{\bm k}^x) - f(\epsilon_{\bm k-\bm q}^y)}{i\omega_n+\epsilon_{\bm k-\bm q}^y-\epsilon_{\bm k}^x} \Bigg\},
\end{align}
where we used $e^{\beta\omega_n}=1$ because of $\omega_n=2n k_B T$.
Then, by the parsing connection, we obtain the imaginary admittance, $\chi^{\pm,\bar\lambda}_{\bm q}(\omega)$ as
\begin{align}
    \chi^{\pm \bar\lambda}_{\bm q}(\omega) 
    \nonumber&= - \sum_{\bm k} \Bigg\{  - \lambda \frac{f(\epsilon_{\bm k-\bm q}^x)-f(\epsilon_{\bm k}^z)}{\hbar\omega+\epsilon_{\bm k-\bm q}^x-\epsilon_{\bm k}^z+i\delta} \mp \frac{f(\epsilon_{\bm k-\bm q}^x)-f(\epsilon_{\bm k}^y)}{\hbar\omega+\epsilon_{\bm k-\bm q}^x-\epsilon_{\bm k}^y+i\delta}  
    \\&\hspace{15mm} - \lambda  \frac{f(\epsilon_{\bm k}^x) - f(\epsilon_{\bm k-\bm q}^z)}{\hbar\omega+\epsilon_{\bm k-\bm q}^z-\epsilon_{\bm k}^x+i\delta} \mp \frac{f(\epsilon_{\bm k}^x) - f(\epsilon_{\bm k-\bm q}^y)}{\hbar\omega+\epsilon_{\bm k-\bm q}^y-\epsilon_{\bm k}^x+i\delta} \Bigg\},
\end{align}
which means
\begin{align}
\nonumber
    \langle L_{\bm q}^{\pm} \rangle (\omega) 
    &= u^\lambda_{\bm q}(\omega) C_{\bm q} 
    \sum_{\bm k} \Bigg\{  -\lambda \frac{f(\epsilon_{\bm k-\bm q}^x)-f(\epsilon_{\bm k}^z)}{\hbar\omega+\epsilon_{\bm k-\bm q}^x-\epsilon_{\bm k}^z+i\delta} \mp \frac{f(\epsilon_{\bm k-\bm q}^x)-f(\epsilon_{\bm k}^y)}{\hbar\omega+\epsilon_{\bm k-\bm q}^x-\epsilon_{\bm k}^y+i\delta}  \\
    &\hspace{25mm} -  \lambda \frac{f(\epsilon_{\bm k}^x) - f(\epsilon_{\bm k-\bm q}^z)}{\hbar\omega+\epsilon_{\bm k-\bm q}^z-\epsilon_{\bm k}^x+i\delta} \mp \frac{f(\epsilon_{\bm k}^x) - f(\epsilon_{\bm k-\bm q}^y)}{\hbar\omega+\epsilon_{\bm k-\bm q}^y-\epsilon_{\bm k}^x+i\delta} \Bigg\}.
\end{align}
When we assume that the input displacement is nearly uniform, the orbital moment becomes 
\begin{align}
    &\langle L_{\bm q}^{\pm} \rangle (\omega) 
    \simeq  2 \hbar\omega u^\lambda_{\bm q}(\omega) C_{\bm q} \sum_{\bm k} \Bigg\{  \lambda \frac{f(\epsilon_{\bm k}^x)-f(\epsilon_{\bm k}^z)}{(\epsilon_{\bm k}^x-\epsilon_{\bm k}^z)^2} \pm \frac{f(\epsilon_{\bm k}^x)-f(\epsilon_{\bm k}^y)}{(\epsilon_{\bm k}^x-\epsilon_{\bm k}^y)^2}  \Bigg\},
\end{align}
and 
\begin{align}
    &\langle L_{\bm q}^{\pm} \rangle (t) 
    \simeq  2 \hbar\omega_{\bm q} u^\lambda_{\bm q}(t) C_{\bm q} \sum_{\bm k} \Bigg\{  \lambda \frac{f(\epsilon_{\bm k}^x)-f(\epsilon_{\bm k}^z)}{(\epsilon_{\bm k}^x-\epsilon_{\bm k}^z)^2} \pm \frac{f(\epsilon_{\bm k}^x)-f(\epsilon_{\bm k}^y)}{(\epsilon_{\bm k}^x-\epsilon_{\bm k}^y)^2}  \Bigg\} ,
\end{align}
which is consistent with the results from the Berry curvature derived in the main article.

The expectation value of the orbital quadrupole is calculated almost in the same manner, and we obtain the response function of the quadrupole, $Q^\pm_{\bm q}$, as
\begin{align}
\nonumber
    \Phi_{\bm q}^{'\pm \bar\lambda} (i\omega_n) 
    &=  \pm \sum_{\bm k} \int_C \frac{dz}{2\pi i} F(z) 
    \\\nonumber
    &\hspace{5mm} \times\Bigg\{ \left(\pm\frac{1}{z-\epsilon_{\bm k}^y} + \lambda \frac{1}{z-\epsilon_{\bm k}^z} \right) \frac{1}{z-i\omega_n-\epsilon_{\bm k-\bm q}^x} 
    + \left(\pm\frac{1}{z-i\omega_n-\epsilon_{\bm k-\bm q}^y}  + \lambda \frac{1}{z-i\omega_n-\epsilon_{\bm k-\bm q}^z} \right)  \frac{1}{z-\epsilon_{\bm k}^x} \Bigg\}
    \\
    &= \mp \sum_{\bm k} \Bigg\{  \lambda \frac{f(\epsilon_{\bm k-\bm q}^x)-f(\epsilon_{\bm k}^z)}{i\omega_n+\epsilon_{\bm k-\bm q}^x-\epsilon_{\bm k}^z} \pm \frac{f(\epsilon_{\bm k-\bm q}^x)-f(\epsilon_{\bm k}^y)}{i\omega_n+\epsilon_{\bm k-\bm q}^x-\epsilon_{\bm k}^y}  
    -  \lambda \frac{f(\epsilon_{\bm k}^x) - f(\epsilon_{\bm k-\bm q}^z)}{i\omega_n+\epsilon_{\bm k-\bm q}^z-\epsilon_{\bm k}^x} \mp \frac{f(\epsilon_{\bm k}^x) - f(\epsilon_{\bm k-\bm q}^y)}{i\omega_n+\epsilon_{\bm k-\bm q}^y-\epsilon_{\bm k}^x} \Bigg\}, \\
    \chi^{'\pm \bar\lambda}_{\bm q}(\omega) 
    &= \mp \sum_{\bm k} \Bigg\{  \lambda \frac{f(\epsilon_{\bm k-\bm q}^x)-f(\epsilon_{\bm k}^z)}{\hbar\omega+\epsilon_{\bm k-\bm q}^x-\epsilon_{\bm k}^z+i\delta} \pm \frac{f(\epsilon_{\bm k-\bm q}^x)-f(\epsilon_{\bm k}^y)}{\hbar\omega+\epsilon_{\bm k-\bm q}^x-\epsilon_{\bm k}^y+i\delta}  \nonumber \\
    &\hspace{10mm} -  \lambda \frac{f(\epsilon_{\bm k}^x) - f(\epsilon_{\bm k-\bm q}^z)}{\hbar\omega+\epsilon_{\bm k-\bm q}^z-\epsilon_{\bm k}^x+i\delta} \mp \frac{f(\epsilon_{\bm k}^x) - f(\epsilon_{\bm k-\bm q}^y)}{\hbar\omega+\epsilon_{\bm k-\bm q}^y-\epsilon_{\bm k}^x+i\delta} \Bigg\},
\end{align}
which leads to
\begin{align}
    &\langle Q_{\bm q}^\pm \rangle (t) 
    \simeq \pm 2 u_{\bm q}^\lambda(t)C_{\bm q}   \sum_{\bm k} \Bigg\{  \lambda \frac{f(\epsilon_{\bm k}^x)-f(\epsilon_{\bm k}^z)}{\epsilon_{\bm k}^x-\epsilon_{\bm k}^z} \pm \frac{f(\epsilon_{\bm k}^x)-f(\epsilon_{\bm k}^y)}{\epsilon_{\bm k}^x-\epsilon_{\bm k}^y}  \Bigg\} ,
\end{align}
with the assumption of $|\bm q| \ll |\bm k|$.

We note that $\langle Q^{yz}_{\bm q}\rangle$ and $\langle Q^{y^2-z^2}_{\bm q}\rangle$ are zero in linear response theory for the same reason as $\langle L^x_{\bm q}\rangle$.

\section{Keldysh Formalism}
\label{app:Keldysh}

The results of $\langle L^x_{\bm q=0}\rangle$ in the main paper can be obtained from the Keldysh formalism.
The entire Hamiltonian reads
\begin{align}
\label{eq:Hamiltonian}
&\mathcal{H}=\mathcal{H}^0_{\rm el} + \mathcal{H}^0_{\rm ph} + \mathcal{H}_{\rm ep}
\\
&\mathcal{H}^0_{\rm el} = \sum_{\bm k ,\alpha=(x,y,z)} \epsilon_{\bm k}^\alpha  c_{\bm k\alpha}^\dag c_{\bm k\alpha},\\
&\mathcal{H}^0_{\rm ph} = \sum_{\bm q,\lambda=(0,\pm)} \hbar\omega_{\bm q} a_{\bm q\lambda}^\dag a_{\bm q\lambda}.
\end{align}
Here, we need to treat the displacement quantum mechanically, and the electron-phonon Hamiltonian for the chiral phonon modes is given as
\begin{align}
    \nonumber\mathcal{H}_{\rm ep} =&  \sum_{\bm k\bm q,\lambda} \sqrt{\frac{\hbar}{2M N\omega_{\bm q\lambda}}} 2i(t_\sigma-t_\pi)( a_{\bm q\lambda} + a^\dag_{-\bm q\bar\lambda}) 
    \\\nonumber&\times\Bigg\{q_x \xi^y_{\bm q\lambda}  \left[c^\dag_{\bm k+\bm q,x} c_{\bm k,y} + c^\dag_{\bm k+\bm q,y} c_{\bm k,x}\right] 
    +q_y \xi^z_{\bm q\lambda}  \left[c^\dag_{\bm k+\bm q,y} c_{\bm k,z} + c^\dag_{\bm k+\bm q,z} c_{\bm k,y}\right]
    + q_x \xi^z_{\bm q\lambda}  \left[ c^\dag_{\bm k+\bm q,z}  c_{\bm k,x} + c^\dag_{\bm k+\bm q,x}  c_{\bm k,z}\right]
    \Bigg\}
    \\\equiv& \sum_{\bm k\bm q} C'_{\bm q} \left[(a_{\bm q,y}+ a_{-\bm q,y}^\dag) ( c_{\bm k+\bm q,x}^\dag  c_{\bm k,y}+ c_{\bm k+\bm q,y}^\dag c_{\bm k,x})  
    + ( a_{\bm q,z}+ a_{-\bm q,z}^\dag)( c_{\bm k+\bm q,z}^\dag  c_{\bm k,x} + c_{\bm k+\bm q,x}^\dag  c_{\bm k,z})\right],
\end{align}
where $C'_{\bm q}=\sqrt{\hbar/2M N\omega_{\bm q}} 2i(t_{\sigma}-t_{\pi})q_x$, which satisfies $(C'_{-\bm q})^*=C'_{\bm q}$.

The net orbital accumulation along $x$-axis is defined by
\begin{align}
    &\langle L^x_{\bm q=0} \rangle 
    \equiv \sum_{\bm k} (c_{\bm k,+}^\dag c_{\bm k,+} - c_{\bm k,-}^\dag c_{\bm k,-}) 
    = 2\, {\rm Im}\, (\sum_{\bm k}  c_{\bm k,y}^\dag c_{\bm k,z}) ,
\end{align}
which is calculated as
\begin{align}
\nonumber
\langle c_{\bm Ky}^\dag(t) c_{\bm Kz}(t)\rangle 
&= \Bigg\langle T_K \Bigg\{e^{-\int dt' \frac{i}{\hbar}\mathcal{H}_{ep}(t') }  c_{\bm Ky}^\dag (t) c_{\bm Kz}(t)\Bigg\} \Bigg\rangle_0 
\\\nonumber&\simeq - \frac{1}{2\hbar^2} \Bigg\langle \int dt' dt'' T_K \Bigg\{\mathcal{H}_{\rm ep}(t')\mathcal{H}_{\rm ep}(t'') c_{\bm Ky}^\dag(t) c_{\bm Kz}(t) \Bigg\}\Bigg\rangle 
\\\nonumber& = - \frac{1}{2\hbar^2}\sum_{\bm k\bm q\gamma\gamma'\lambda} \sum_{\bm k'\bm q'\mu\mu'\lambda'} 
\Bigg\langle \int dt' dt'' T_K \Bigg\{C'_{\bm q}C'_{\bm q'} (a_{\bm q\lambda}(t')+a_{-\bm q\lambda}^\dag (t')) (a_{\bm q'\lambda'}(t'')+a_{-\bm q'\lambda'}^\dag (t''))
\\&\hspace{2cm}\times  c_{\bm k+\bm q\gamma'}^\dag(t') c_{\bm k\gamma}(t') c_{\bm k'+\bm q'\mu'}^\dag(t'') c_{\bm k'\mu}(t'') c_{\bm Ky}^\dag(t) c_{\bm Kz}(t) \Bigg\}\Bigg\rangle ,
\end{align}
where $T_K$ is the time-ordering operator along the Keldysh contour.
By picking up the connected diagram, we obtain
\begin{align}
\langle c_{\bm Ky}^\dag(t) c_{\bm Kz}(t)\rangle 
\nonumber&= \frac{1}{\hbar^2}\sum_{\bm k} |C'_{\bm K-\bm k}|^2 
\int dt' dt''   \langle T_K  c_{\bm Kz}(t) c_{\bm Kz}^\dag(t') \rangle \langle T_K c_{\bm kx}(t') c_{\bm kx}^\dag(t'')\rangle \langle T_K c_{\bm Ky}(t'') c_{\bm Ky}^\dag(t) \rangle 
\\&\hspace{5mm}\times\Bigg\{ \langle T_K a_{\bm K-\bm k,z}(t') a_{\bm K-\bm k,y}^\dag (t'')\rangle + \langle T_K  a_{\bm k-\bm K,y}(t'') a_{\bm k-\bm K,z}^\dag(t') \rangle \Bigg\}.
\end{align}
Following the Keldysh formalism, we define the Green's function as
\begin{align}
G^T(1,2) &\equiv -i \langle \mathcal{T} \psi(1) \psi^\dag(2) \rangle,\\
G^>(1,2) &\equiv -i\langle \psi(1) \psi^\dag(2)\rangle,\\
G^<(1,2) &\equiv \pm i \langle  \psi^\dag(2) \psi(1)\rangle,\\
G^{\bar{T}}(1,2) &\equiv -i\langle \bar{\mathcal{T}} \psi(1) \psi^\dag(2)\rangle,
\end{align}
where the upper sign denotes the Fermion, and $\mathcal{T}$ is the time-ordering operator: $\mathcal{T} A(t) B(t')=\Theta(t-t') A(t) B(t')\mp \Theta(t'-t) B(t') A(t)$.
Then, the above equation is transformed as
\begin{align}
\nonumber
\langle c_{\bm Ky}^\dag(t) c_{\bm Kz}(t)\rangle &= \frac{1}{\hbar^2}\sum_{\bm k}   |C'_{\bm K-\bm k}|^2  \int_{-\infty}^\infty dt' dt''
\\\nonumber& \hspace{5mm} \times \Bigg\{ G_{\bm Kz}^T(t,t') G_{\bm k x}^T(t',t'') G_{\bm Ky}^<(t'',t) 
\left[D_{\bm K-\bm k}^{zy,T}(t',t'') + D_{\bm k-\bm K}^{yz,T}(t'',t')\right]
\\\nonumber&\hspace{1cm} -G_{\bm Kz}^T(t,t') G_{\bm k x}^<(t',t'') G_{\bm Ky}^{\bar{T}}(t'',t) 
\left[D_{\bm K-\bm k}^{zy,<}(t',t'') + D_{\bm k-\bm K}^{yz,>}(t'',t')\right] 
\\\nonumber&\hspace{1cm} -G_{\bm Kz}^<(t,t') G_{\bm k x}^>(t',t'') G_{\bm Ky}^<(t'',t) 
\left[D_{\bm K-\bm k}^{zy,>}(t',t'') + D_{\bm k-\bm K}^{yz,<}(t'',t')\right] 
\\\nonumber&\hspace{1cm} +G_{\bm Kz}^<(t,t') G_{\bm k x}^{\bar{T}}(t',t'') G_{\bm Ky}^{\bar{T}}(t'',t) 
\left[D_{\bm K-\bm k}^{zy,\bar{T}}(t',t'') + D_{\bm k-\bm K}^{yz,\bar{T}}(t'',t')\right] \Bigg\}
\\\nonumber&= \frac{1}{\hbar^2} \sum_{\bm k}  |C'_{\bm K-\bm k}|^2  \int_{-\infty}^\infty \frac{d\omega d\omega'}{(2\pi)^2}
\\\nonumber&  \hspace{5mm}  \times \Bigg\{ G_{\bm Kz}^T(\omega) G_{\bm k x}^T(\omega') G_{\bm Ky}^<(\omega) 
\left[D_{\bm K-\bm k}^{zy,T}(\omega-\omega') + D_{\bm k-\bm K}^{yz,T}(\omega'-\omega)\right]
\\\nonumber&\hspace{1cm} -G_{\bm Kz}^T(\omega) G_{\bm k x}^<(\omega') G_{\bm Ky}^{\bar{T}}(\omega) 
\left[D_{\bm K-\bm k}^{zy,<}(\omega-\omega') + D_{\bm k-\bm K}^{yz,>}(\omega'-\omega)\right] 
\\\nonumber&\hspace{1cm} -G_{\bm Kz}^<(\omega) G_{\bm k x}^>(\omega') G_{\bm Ky}^<(\omega) 
\left[D_{\bm K-\bm k}^{zy,>}(\omega-\omega') + D_{\bm k-\bm K}^{yz,<}(\omega'-\omega)\right] 
\\&\hspace{1cm} +G_{\bm Kz}^<(\omega) G_{\bm k x}^{\bar{T}}(\omega') G_{\bm Ky}^{\bar{T}}(\omega) 
\left[D_{\bm K-\bm k}^{zy,\bar{T}}(\omega-\omega') + D_{\bm k-\bm K}^{yz,\bar{T}}(\omega'-\omega)\right] \Bigg\},
\end{align}
where $G(\omega) \equiv \int_{-\infty}^\infty dt e^{i\omega t} G(t,0)$, and the negative sign in the second and third terms comes from the direction of the Keldysh path.
Since the Keldysh Green's functions follow the relations of
\begin{align}
G^<_{\bm k}(\omega) &= \mp 2i n_{\bm k} \, {\rm Im}\, G^R_{\bm k}(\omega),\\
G^>_{\bm k}(\omega) &= 2i (1\mp n_{\bm k})\, {\rm Im}\, G^R_{\bm k}(\omega),\\
G^T &=G^R+G^<,\\
G^{\bar{T}} &= G^>-G^R,\\
{\rm Re}\, G^T_{\bm k}(\omega) &= - \, {\rm Re}\, G^{\bar{T}}_{\bm k}(\omega) = {\rm Re}\, G^R_{\bm k}(\omega),\\
{\rm Im}\, G^T_{\bm k}(\omega) &= {\rm Im}\, G^{\bar{T}}_{\bm k}(\omega) = (1\mp 2n_{\bm k})\, {\rm Im}\, G^R_{\bm k}(\omega).
\end{align}
In particular, using $(G^T)^*=-G^{\bar{T}}$ and ${\rm Re} \, G^<={\rm Re}\, G^>=0$, and noting that $y$ and $z$ in $D_{\bm K-\bm k}$ also interchange under complex conjugation, we obtain 
\begin{align}
\nonumber
{\rm Im} \, \langle c_{\bm Ky}^\dag(t) c_{\bm Kz}(t)\rangle 
& =\frac{1}{2i\hbar^2} \sum_{\bm k}  |C'_{\bm K-\bm k}|^2  \int_{-\infty}^\infty \frac{d\omega d\omega'}{(2\pi)^2}
\\\nonumber& \hspace{5mm} \times \Bigg\{ G_{\bm Kz}^T(\omega) G_{\bm k x}^T(\omega') G_{\bm Ky}^<(\omega) 
\left[D_{\bm K-\bm k}^{zy,T}(\omega-\omega') + D_{\bm k-\bm K}^{yz,T}(\omega'-\omega)\right]
\\\nonumber&\hspace{1cm} -G_{\bm Kz}^T(\omega) G_{\bm k x}^<(\omega') G_{\bm Ky}^{\bar{T}}(\omega) 
\left[D_{\bm K-\bm k}^{zy,<}(\omega-\omega') + D_{\bm k-\bm K}^{yz,>}(\omega'-\omega)\right] 
\\\nonumber&\hspace{1cm} +G_{\bm Kz}^<(\omega) G_{\bm k x}^{\bar{T}}(\omega') G_{\bm Ky}^{\bar{T}}(\omega) 
\left[D_{\bm K-\bm k}^{zy,\bar{T}}(\omega-\omega') + D_{\bm k-\bm K}^{yz,\bar{T}}(\omega'-\omega)\right]
\\\nonumber&\hspace{1cm}-G_{\bm Kz}^{\bar{T}}(\omega) G_{\bm k x}^{\bar{T}}(\omega') G_{\bm Ky}^<(\omega) 
\left[D_{\bm K-\bm k}^{yz,\bar T}(\omega-\omega') + D_{\bm k-\bm K}^{zy,\bar T}(\omega'-\omega)\right]
\\\nonumber&\hspace{1cm} +G_{\bm Kz}^{\bar T}(\omega) G_{\bm k x}^<(\omega') G_{\bm Ky}^{T}(\omega) 
\left[D_{\bm K-\bm k}^{yz,<}(\omega-\omega') + D_{\bm k-\bm K}^{zy,>}(\omega'-\omega)\right] 
\\&\hspace{1cm} -G_{\bm Kz}^<(\omega) G_{\bm k x}^{T}(\omega') G_{\bm Ky}^{T}(\omega) 
\left[D_{\bm K-\bm k}^{yz,T}(\omega-\omega') + D_{\bm k-\bm K}^{zy,T}(\omega'-\omega)\right]\Bigg\}.
\label{eq:Imcc}
\end{align}

So far, we derived the orbital moment in general, but hereafter, we consider injecting chiral modes described by
\begin{align}
    \langle a_{\bm q+}^\dag a_{\bm q+}\rangle = n_{\bm q}^+,\,\,\,
    \langle a_{\bm q-}^\dag a_{\bm q-}\rangle = n_{\bm q}^-,
\end{align}
where $a_{\bm q,\pm} = \frac{1}{\sqrt{2}} (a_{\bm q,y} \pm i a_{\bm q,z})$.
Considering 
\begin{align}
    & a_{\bm q,y} a_{\bm q,z}^\dag = - \frac{1}{2i}(a_{\bm q,+}+a_{\bm q,-})(a_{\bm q,+}^\dag - a_{\bm q,-}^\dag),  \,\,\,
    \\
    & a_{\bm q,z} a_{\bm q,y}^\dag= \frac{1}{2i}(a_{\bm q,+} - a_{\bm q,-})(a_{\bm q,+}^\dag + a_{\bm q,-}^\dag),
\end{align}
and neglecting $\langle a_\pm^\dag a_\mp\rangle$, the Green's function of phonons are written as
\begin{align}
    &D^{yz,\zeta}_{\bm q}(\omega) = - \frac{1}{2i} \left[D^{+,\zeta}_{\bm q}(\omega) - D^{-,\zeta}_{\bm q}(\omega)\right],\,\,\,
    \\
    &D^{zy,\zeta}_{\bm q}(\omega) =  \frac{1}{2i} \left[D^{+,\zeta}_{\bm q}(\omega) - D^{-,\zeta}_{\bm q}(\omega)\right],
\end{align}
where we defined $D^\pm_{\bm q}= D^{\pm\pm}_{\bm q}$.

Then, returning to the calculation of Eq.~(\ref{eq:Imcc}), and recalling that the net orbital accumulation is defined from $\langle L^x_{\bm q=0}\rangle = 2\sum_k {\rm Im}\, \langle c_{\bm k,y}^\dag c_{\bm k,z}\rangle $, we obtain
\begin{align}
\langle L^x_{\bm q=0} \rangle
\nonumber&= - \frac{1}{2\hbar^2} \sum_{\bm k,\bm K,\lambda} \lambda |C'_{\bm K-\bm k}|^2  \int_{-\infty}^\infty \frac{d\omega d\omega'}{(2\pi)^2}
\\\nonumber& \hspace{5mm} \times \Bigg\{ G_{\bm Kz}^T(\omega) G_{\bm k x}^T(\omega') G_{\bm Ky}^<(\omega) 
\left[D_{\bm K-\bm k}^{\lambda,T}(\omega-\omega') - D_{\bm k-\bm K}^{\lambda,T}(\omega'-\omega)\right]
\\\nonumber&\hspace{1cm} -G_{\bm Kz}^T(\omega) G_{\bm k x}^<(\omega') G_{\bm Ky}^{\bar{T}}(\omega) 
\left[D_{\bm K-\bm k}^{\lambda,<}(\omega-\omega') - D_{\bm k-\bm K}^{\lambda,>}(\omega'-\omega)\right] 
\\\nonumber&\hspace{1cm} +G_{\bm Kz}^<(\omega) G_{\bm k x}^{\bar{T}}(\omega') G_{\bm Ky}^{\bar{T}}(\omega) 
\left[D_{\bm K-\bm k}^{\lambda,\bar{T}}(\omega-\omega') - D_{\bm k-\bm K}^{\lambda,\bar{T}}(\omega'-\omega)\right]
\\\nonumber&\hspace{1cm}-G_{\bm Kz}^{\bar{T}}(\omega) G_{\bm k x}^{\bar{T}}(\omega') G_{\bm Ky}^<(\omega) 
\left[-D_{\bm K-\bm k}^{\lambda,\bar T}(\omega-\omega') + D_{\bm k-\bm K}^{\lambda,\bar T}(\omega'-\omega)\right]
\\\nonumber&\hspace{1cm} +G_{\bm Kz}^{\bar T}(\omega) G_{\bm k x}^<(\omega') G_{\bm Ky}^{T}(\omega)
\left[-D_{\bm K-\bm k}^{\lambda,<}(\omega-\omega') + D_{\bm k-\bm K}^{\lambda,>}(\omega'-\omega)\right] 
\\&\hspace{1cm} -G_{\bm Kz}^<(\omega) G_{\bm k x}^{T}(\omega') G_{\bm Ky}^{T}(\omega) 
\left[-D_{\bm K-\bm k}^{\lambda,T}(\omega-\omega') + D_{\bm k-\bm K}^{\lambda,T}(\omega'-\omega)\right]\Bigg\}
\end{align}
When we assume that the phonons have no relaxation and can be described by the bare Green's functions given by
\begin{align}
    &D_{\bm q}^T(\omega) = \frac{1}{\omega-\omega_{\bm q}+i0} - 2\pi in_{\bm q}\delta(\omega-\omega_{\bm q}),\,\,\,
    \\
    &D_{\bm q}^<(\omega) = - 2\pi in_{\bm q}\delta(\omega-\omega_{\bm q}),\,\,\,
    \\
    &D_{\bm q}^>(\omega) = - 2\pi i(1+n_{\bm q})\delta(\omega-\omega_{\bm q}),
\end{align}
we can derive
\begin{align}
&D_{\bm q}^{\lambda,T}(\omega) - D_{-\bm q}^{\lambda,T}(-\omega) 
= \frac{2\omega}{\omega^2-\omega_{\bm q}^2} - 2\pi i \left[ n_{\bm q}^\lambda\delta(\omega-\omega_{\bm q}) -n_{-\bm q}^\lambda\delta(-\omega-\omega_{\bm q})\right],
\\
&D_{\bm q}^{\lambda,<}(\omega) - D_{-\bm q}^{\lambda,>}(-\omega) 
= - 2\pi i \left[ n_{\bm q}^\lambda\delta(\omega-\omega_{\bm q}) -(1+n_{-\bm q}^\lambda)\delta(-\omega-\omega_{\bm q})\right],
\end{align}
because of $\omega_{\bm q}=\omega_{-\bm q}$.
Since the terms without $\lambda$ dependence disappears by the summation of $\lambda$, we can replace as
\begin{align}
    &D_{\bm q}^{\lambda,T}(\omega) - D_{-\bm q}^{\lambda,T}(-\omega) 
    \rightarrow - 2\pi i \left[ n_{\bm q}^\lambda\delta(\omega-\omega_{\bm q}) -n_{-\bm q}^\lambda\delta(-\omega-\omega_{\bm q})\right],
    \\
    &D_{\bm q}^{\lambda,<}(\omega) - D_{-\bm q}^{\lambda,>}(-\omega)
    \rightarrow - 2\pi i \left[ n_{\bm q}^\lambda\delta(\omega-\omega_{\bm q}) - n_{-\bm q}^\lambda \delta(-\omega-\omega_{\bm q})\right],
\end{align}
which are completely the same.
Thus, we obtain
\begin{align}
\nonumber
\langle L^x_{\bm 0} \rangle
&= - \frac{1}{2\hbar^2} \sum_{\bm k,\bm K,\lambda} \lambda |C'_{\bm K-\bm k}|^2  \int_{-\infty}^\infty \frac{d\omega d\omega'}{2\pi} 
\left[ n_{\bm K-\bm k}^\lambda\delta(\omega-\omega'-\omega_{\bm K-\bm k}) - n_{\bm k-\bm K}^\lambda \delta(\omega'-\omega-\omega_{\bm k-\bm K})\right]
\\\nonumber& \hspace{5mm} \times 2 \, {\rm Im}\, \Bigg\{  G_{\bm k x}^T(\omega') \left[G_{\bm Kz}^T(\omega)  G_{\bm Ky}^<(\omega) + G_{\bm Ky}^T(\omega)  G_{\bm Kz}^<(\omega)\right] 
- G_{\bm k x}^<(\omega') G_{\bm Kz}^T(\omega)  G_{\bm Ky}^{\bar{T}}(\omega) \Bigg\}
\\\nonumber&= - \frac{1}{2\hbar^2} \sum_{\bm k,\bm q} |C'_{\bm q}|^2  \int_{-\infty}^\infty \frac{d\omega d\omega'}{2\pi} (n_{\bm q}^+ - n_{\bm q}^-) 
\left[ \delta(\omega-\omega'-\omega_{\bm q}) - \delta(\omega'-\omega-\omega_{\bm q})\right]
\\&  \hspace{5mm}  \times 2 \, {\rm Im}\, \Bigg\{  - G_{\bm k x}^<(\omega') G_{\bm k+\bm q,z}^T(\omega)  G_{\bm k+\bm q,y}^{\bar{T}}(\omega) 
+ G_{\bm k x}^T(\omega') \left[G_{\bm k+\bm q,z}^T(\omega)  G_{\bm k+\bm q,y}^<(\omega) + G_{\bm k+\bm q,y}^T(\omega)  G_{\bm k+\bm q,z}^<(\omega)\right] \Bigg\},
\end{align}
since the electronic Green's function satisfies $G_{\bm k}=G_{-\bm k}$ in our inversion symmetric model.
We note that it is already clear at this point that the time reversal symmetry breaking is crucial for the generation of the net orbital accumulation.

For the free electrons, the Green's functions of the electrons are given as
\begin{align}
&G_{\bm kx}^T(\omega) = \frac{1}{\omega-\epsilon_{\bm kx}/\hbar+i0} + 2\pi if_{\bm k}^x \delta(\omega-\epsilon_{\bm kx}/\hbar),
\\
&G_{\bm ky}^T(\omega) = \frac{1}{\omega-\epsilon_{\bm ky}/\hbar+i0} + 2\pi if_{\bm k}^y \delta(\omega-\epsilon_{\bm ky}/\hbar),
\\
&G_{\bm kz}^T(\omega) = \frac{1}{\omega-\epsilon_{\bm kz}/\hbar+i0} + 2\pi if_{\bm k}^z \delta(\omega-\epsilon_{\bm kz}/\hbar),
\\
&G_{\bm ky}^{\bar T}(\omega) = -\frac{1}{\omega-\epsilon_{\bm ky}/\hbar-i0} + 2\pi if_{\bm k}^y \delta(\omega-\epsilon_{\bm ky}/\hbar),
\\
&G_{\bm kz}^{\bar T}(\omega) = -\frac{1}{\omega-\epsilon_{\bm kz}/\hbar-i0} + 2\pi if_{\bm k}^z \delta(\omega-\epsilon_{\bm kz}/\hbar),
\\
&G_{\bm ky}^<(\omega) = 2\pi if_{\bm k}^y \delta(\omega-\epsilon_{\bm ky}/\hbar),
\\
&G_{\bm kz}^<(\omega) = 2\pi if_{\bm k}^z \delta(\omega-\epsilon_{\bm kz}/\hbar).
\end{align}
When we assume that the energy bands are non-degenerate, the terms containing two delta functions vanish, and we obtain
\begin{align}
\nonumber
\langle L^x_{\bm 0} \rangle 
&=   -\sum_{\bm k,\bm q} |C'_{\bm q}|^2   (n_{\bm q}^+ - n_{\bm q}^- ) \int d\omega
\\\nonumber& \hspace{5mm} \times \Bigg\{  \Bigg[ \frac{1}{\hbar\omega-\epsilon_{\bm k+\bm q}^z} f_{\bm k+\bm q}^y\delta(\omega-\epsilon_{\bm k+\bm q}^y/\hbar)  
+ \frac{1}{\hbar\omega-\epsilon_{\bm k+\bm q}^y} f_{\bm k+\bm q}^z\delta(\omega-\epsilon_{\bm k+\bm q}^z/\hbar) \Bigg] 
\left(\frac{1}{\hbar\omega-\hbar\omega_{\bm q}-\epsilon_{\bm k}^x} - \frac{1}{\hbar\omega+\hbar\omega_{\bm q}-\epsilon_{\bm k}^x}\right)
\\\nonumber&\hspace{1cm} + f_{\bm k}^x\left[\delta(\omega-\omega_{\bm q}-\epsilon_{\bm k}^x/\hbar) - \delta(\omega+\omega_{\bm q}-\epsilon_{\bm k}^x/\hbar) \right] \frac{1}{\hbar\omega-\epsilon_{\bm k+\bm q}^z}\frac{1}{\hbar\omega-\epsilon_{\bm k+\bm q}^y} \Bigg\}
\\\nonumber&= -\sum_{\bm k,\bm q} |C'_{\bm q}|^2   (n_{\bm q}^+ - n_{\bm q}^- ) 
\\\nonumber& \hspace{5mm} \times \Bigg\{  \frac{f_{\bm k}^y }{\epsilon_{\bm k}^y-\epsilon_{\bm k}^z} \left(\frac{1}{\epsilon_{\bm k}^y-\epsilon_{\bm k-\bm q}^x-\hbar\omega_{\bm q}} - \frac{1}{\epsilon_{\bm k}^y-\epsilon_{\bm k-\bm q}^x+\hbar\omega_{\bm q}}\right) 
+ \frac{f_{\bm k}^z }{\epsilon_{\bm k}^z-\epsilon_{\bm k}^y} \left(\frac{1}{\epsilon_{\bm k}^z-\epsilon_{\bm k-\bm q}^x-\hbar\omega_{\bm q}} - \frac{1}{\epsilon_{\bm k}^z-\epsilon_{\bm k-\bm q}^x+\hbar\omega_{\bm q}}\right)
\\&\hspace{1cm} + f_{\bm k}^x \Bigg[ \frac{1}{\epsilon_{\bm k}^x-\epsilon_{\bm k+\bm q}^z+\hbar\omega_{\bm q}}\frac{1}{\epsilon_{\bm k}^x-\epsilon_{\bm k+\bm q}^y+\hbar\omega_{\bm q}} 
- \frac{1}{\epsilon_{\bm k}^x-\epsilon_{\bm k+\bm q}^z-\hbar\omega_{\bm q}}\frac{1}{\epsilon_{\bm k}^x-\epsilon_{\bm k+\bm q}^y-\hbar\omega_{\bm q}} \Bigg] \Bigg\},
\label{eq:Keldysh_result}
\end{align}
which is consistent with the derivation based on the Berry curvature given in the main article in the long-wavelength and low-frequency limit of phonons.
We note that although the zeroth order terms about $\hbar\omega_{\bm q}$ are finite, it vanishes when we assume $|\bm q|\ll|\bm k|$.

In the numerical evaluation, the degeneracy is regularized by a finite broadening parameter $\Gamma$.
In the cusp region in Fig.~4(a) of the main text, where the $p_x$, $p_y$, and $p_z$ bands come close to each other near the Fermi surface, the magnitude of the orbital accumulation becomes strongly dependent on the broadening factor $\Gamma$.

\section{Origin of the Orbital Accumulation}
\label{sec:physical_interpretation}

In this section, we explicitly show that the orbital accumulation, $\langle L^x_{\bm q=0}\rangle$, is induced by the modification of the wavefunctions for the occupied states.
By second-order perturbation with respect to ${\cal H}_{\rm ep}$, the wavefunction is written as
\begin{align}
    \ket{\psi_{\alpha}} &= \ket{\bm k \alpha}\!\ket{\rm ph}+ \sum_{\bm k' \beta} \frac{\bra{\bm k' \beta}\bra{{\rm ph}_1}\mathcal{H}_{\rm ep}\ket{\bm k \alpha}\!\ket{\rm ph}}{\epsilon_{\bm k}^{\alpha}-\epsilon_{\bm k'}^{\beta}+E_{\rm ph}-E_{{\rm ph}_1}} \ket{\bm k' \beta}\! \ket{{\rm ph}_1}
    \nonumber\\\nonumber
    &-\frac{1}{2}\sum_{\bm k'\beta} \frac{\bra{\bm k \alpha}\bra{\rm ph}\mathcal{H}_{\rm ep}\ket{\bm k' \beta}\!\ket{{\rm ph}_1}\bra{\bm k'\beta}\!\bra{{\rm ph}_1}\mathcal{H}_{\rm ep}\ket{\bm k \beta}\!\ket{\rm ph}}{(\epsilon_{\bm k}^{\alpha} - \epsilon_{\bm k'}^{\beta}+E_{\rm ph}-E_{{\rm ph}_1})^2} \ket{\bm k \alpha}\! \ket{\rm ph}
    \\\nonumber
    &+ \sum_{\bm k' \beta} \Bigg(\sum_{\bm k''\gamma\lambda'
    } \frac{\bra{\bm k'\beta}\!\bra{{\rm ph}_1}\mathcal{H}_{\rm ep}\ket{\bm k''\gamma}\!\ket{{\rm ph}_2}\bra{\bm k''\gamma}\!\bra{{\rm ph}_2}\mathcal{H}_{\rm ep}\ket{\bm k \alpha}\!\ket{\rm ph}}{(\epsilon_{\bm k}^{\alpha}-\epsilon_{\bm k''}^{\gamma}+E_{\rm ph}-E_{{\rm ph}_2})(\epsilon_{\bm k}^{\alpha}-\epsilon_{\bm k'}^{\beta} +E_{\rm ph}-E_{{\rm ph}_1})} \ket{\bm k' \beta}\!\ket{{\rm ph}_1}
    \\&\hspace{5mm}- \frac{\bra{\bm k \alpha}\!\bra{\rm ph}\mathcal{H}_{\rm ep}\ket{\bm k \alpha}\!\ket{\rm ph}\bra{\bm k'\beta}\!\bra{{\rm ph}_1}\mathcal{H}_{\rm ep}\ket{\bm k \alpha}\!\ket{\rm ph}}{(\epsilon_{\bm k}^{\alpha} - \epsilon_{\bm k'}^{\beta}+E_{\rm ph}-E_{{\rm ph}_1})^2} \ket{\bm k'\beta}\!\ket{{\rm ph}_1}\Bigg) , 
\end{align}
where $\ket{\rm ph}$ is the state of phonons, and $E_{\rm ph}$ is the total energy of the phonon bath.
If the orbital accumulation stems from the entire Fermi sea, it is given as
\begin{align}
\langle  L^x \rangle = \sum_{{\bm k},\alpha} f_{\bm k}^{\alpha} n_{\rm ph} \bra{\psi_{\alpha}}  L^x \ket{\psi_{\alpha}},
\end{align}
where $f_{\bm k}^\alpha$ and $n_{\rm ph}$ are the population of electrons and phonons, respectively.
For example, the orbital accumulation originating from the perturbed state of $\ket{\bm k y}\ket{\rm ph}$, which have the population of $f_{\bm k}^y n_{\rm ph}$, is given as
\begin{align}
    &f_{\bm k}^y n_{\rm ph} \bra{\psi_{y}} {L}^x \ket{\psi_{y}}\nonumber 
    \\\nonumber&= 2 f_{\bm k}^y n_{\rm ph} \, {\rm Im}\, \Bigg\{ \sum_{\bm q} \frac{\bra{\bm k z}\bra{\rm ph}\mathcal{H}_{\rm ep}\ket{\bm k\pm\bm q x}\ket{{\rm ph}_1}\bra{\bm k\pm\bm q x}\bra{{\rm ph}_1}\mathcal{H}_{\rm ep}\ket{\bm k y}\ket{\rm ph}}{(\epsilon_{\bm k}^y-\epsilon_{\bm k\pm\bm q}^x+E_{\rm ph}-E_{{\rm ph}_1})(\epsilon_{\bm k}^y-\epsilon_{\bm k}^z)} \Bigg\}
    \\\nonumber&= \sum_{\bm q} 2\, {\rm Im}\, \Bigg\{ \frac{|C'_{\bm q}|^2\langle a_{\bm q,z}^\dag a_{\bm q,y}\rangle f_{\bm k}^y}{(\epsilon_{\bm k}^y-\epsilon_{\bm k+\bm q}^x+\hbar\omega_{\bm q})(\epsilon_{\bm k}^y-\epsilon_{\bm k}^z)} 
    +   \frac{|C'_{\bm q}|^2 \langle a_{-\bm q,z} a_{-\bm q,y}^\dag\rangle f_{\bm k}^y}{(\epsilon_{\bm k}^y-\epsilon_{\bm k+\bm q}^x-\hbar\omega_{\bm q})(\epsilon_{\bm k}^y-\epsilon_{\bm k}^z)} \Bigg\}
    \\&= -\sum_{\bm q} |C'_{\bm q}|^2 (n_{\bm q}^+ - n_{\bm q}^-) f_{\bm k}^y  
     \Bigg\{ \frac{1}{(\epsilon_{\bm k}^y-\epsilon_{\bm k+\bm q}^x-\hbar\omega_{\bm q})(\epsilon_{\bm k}^y-\epsilon_{\bm k}^z)} 
    - \frac{1}{(\epsilon_{\bm k}^y-\epsilon_{\bm k-\bm q}^x+\hbar\omega_{\bm q})(\epsilon_{\bm k}^y-\epsilon_{\bm k}^z)} \Bigg\},
\end{align}
where we used $\langle a_{\bm q,z}^\dag a_{\bm q,y}\rangle = -\langle a_{\bm q,y}^\dag a_{\bm q,z}\rangle = -\frac{1}{2i} \left\{\langle a_{\bm q,+}^\dag a_{\bm q,+}\rangle - \langle a_{\bm q,-}^\dag a_{\bm q,-}\rangle \right\}$ in the chiral mode. 
Also, the accumulation originating from $\ket{\bm k x}\ket{\rm ph}$ reads
\begin{align}
f_{\bm k}^x n_{\rm ph} \bra{\psi_x}{L}^x\ket{\psi_x}
&= -f_{\bm k}^x \sum_{\bm q}|C'_{\bm q}|^2 (n_{\bm q}^+ - n_{\bm q}^-) \nonumber
\\&\hspace{5mm}\times\Bigg\{ \frac{1}{(\epsilon_{\bm k}^x-\epsilon_{\bm k+\bm q}^z+\hbar\omega_{\bm q})(\epsilon_{\bm k}^x-\epsilon_{\bm k+\bm q}^y+\hbar\omega_{\bm q})} 
- \frac{1}{(\epsilon_{\bm k}^x-\epsilon_{\bm k+\bm q}^z-\hbar\omega_{\bm q})(\epsilon_{\bm k}^x-\epsilon_{\bm k+\bm q}^y-\hbar\omega_{\bm q})} \Bigg\}.
\end{align}
Finally, by taking the summation of $\bm k$, we obtain the net orbital accumulation as
\begin{align}
\nonumber
\langle L^x_{\bm q=0} \rangle
&= -\sum_{\bm k} \sum_{\bm q}  |C'_{\bm q}|^2 (n_{\bm q}^+ - n_{\bm q}^-) 
\\\nonumber&\hspace{1cm}\times\Bigg\{ \frac{f_{\bm k}^y}{\epsilon_{\bm k}^y-\epsilon_{\bm k}^z} \Bigg[ \frac{1}{\epsilon_{\bm k}^y-\epsilon_{\bm k+\bm q}^x-\hbar\omega_{\bm q}} -   \frac{1}{\epsilon_{\bm k}^y-\epsilon_{\bm k-\bm q}^x+\hbar\omega_{\bm q}}\Bigg] 
+ \frac{f_{\bm k}^z}{\epsilon_{\bm k}^z-\epsilon_{\bm k}^y} \Bigg[ \frac{1}{\epsilon_{\bm k}^z-\epsilon_{\bm k+\bm q}^x-\hbar\omega_{\bm q}} -   \frac{1}{\epsilon_{\bm k}^z-\epsilon_{\bm k-\bm q}^x+\hbar\omega_{\bm q}}\Bigg]
\\\nonumber&\hspace{2cm} + f_{\bm k}^x \Bigg[ \frac{1}{(\epsilon_{\bm k}^x-\epsilon_{\bm k+\bm q}^z+\hbar\omega_{\bm q})(\epsilon_{\bm k}^x-\epsilon_{\bm k+\bm q}^y+\hbar\omega_{\bm q})} - \frac{1}{(\epsilon_{\bm k}^x-\epsilon_{\bm k+\bm q}^z-\hbar\omega_{\bm q})(\epsilon_{\bm k}^x-\epsilon_{\bm k+\bm q}^y-\hbar\omega_{\bm q})} \Bigg] \Bigg\}
\\\nonumber&\simeq -\sum_{\bm k} \sum_{\bm q}  |C'_{\bm q}|^2 (n_{\bm q}^+ - n_{\bm q}^-) 
\\\nonumber&\hspace{1cm}\times\Bigg\{ \frac{f_{\bm k}^y}{\epsilon_{\bm k}^y-\epsilon_{\bm k}^z} \Bigg[ \frac{1}{\epsilon_{\bm k}^y-\epsilon_{\bm k}^x-\hbar\omega_{\bm q}} -   \frac{1}{\epsilon_{\bm k}^y-\epsilon_{\bm k}^x+\hbar\omega_{\bm q}}\Bigg] 
+ \frac{f_{\bm k}^z}{\epsilon_{\bm k}^z-\epsilon_{\bm k}^y} \Bigg[ \frac{1}{\epsilon_{\bm k}^z-\epsilon_{\bm k}^x-\hbar\omega_{\bm q}} -   \frac{1}{\epsilon_{\bm k}^z-\epsilon_{\bm k}^x+\hbar\omega_{\bm q}}\Bigg]
\\\nonumber&\hspace{2cm} + f_{\bm k}^x \Bigg[ \frac{1}{(\epsilon_{\bm k}^x-\epsilon_{\bm k}^z+\hbar\omega_{\bm q})(\epsilon_{\bm k}^x-\epsilon_{\bm k}^y+\hbar\omega_{\bm q})} - \frac{1}{(\epsilon_{\bm k}^x-\epsilon_{\bm k}^z-\hbar\omega_{\bm q})(\epsilon_{\bm k}^x-\epsilon_{\bm k}^y-\hbar\omega_{\bm q})} \Bigg] \Bigg\}
\\\nonumber&\simeq -\sum_{\bm k} \sum_{\bm q}  2|C'_{\bm q}|^2 (n_{\bm q}^+ - n_{\bm q}^-) \hbar\omega_{\bm q}
\\&\hspace{3mm}\times\Bigg\{ \frac{f_{\bm k}^y}{(\epsilon_{\bm k}^y-\epsilon_{\bm k}^z )(\epsilon_{\bm k}^y-\epsilon_{\bm k}^x)^2}  + \frac{f_{\bm k}^z}{(\epsilon_{\bm k}^z-\epsilon_{\bm k}^y )(\epsilon_{\bm k}^z-\epsilon_{\bm k}^x)^2 } 
- f_{\bm k}^x \Bigg[ \frac{1}{(\epsilon_{\bm k}^x-\epsilon_{\bm k}^z)^2(\epsilon_{\bm k}^x-\epsilon_{\bm k}^y)} + \frac{1}{(\epsilon_{\bm k}^x-\epsilon_{\bm k}^z)(\epsilon_{\bm k}^x-\epsilon_{\bm k}^y)^2} \Bigg] \Bigg\},
\end{align}
which is consistent with the results of the Keldysh formalism.
This result proves that the orbital accumulation stems from the entire Fermi sea, i.e., the sum of the modified eigenstates under second-order perturbation due to electron-phonon coupling.

\section{Equations of motion}

The equations of motion given in Eqs.~(26)-(29) in the main text can be solved as
\begin{align}
\langle  L^x_{\bm q} \rangle &= \langle  Q^{yz}_{\bm q} \rangle = 0. \label{Lxsol} \\
\langle  L_{\bm q}^y\rangle &= \sum_{\bm k} \frac{2i\lambda C_{\bm q}\dot{u}_{\bm q}^\lambda (f_{\bm k}^z-f_{\bm k}^x)}{\hbar^2\omega_{\bm q}^2 - (\epsilon^x_{\bm k}-\epsilon^z_{\bm k})^2} ,\label{Lysol} \\
\langle  Q^{zx}_{\bm q,\bm k} \rangle &= \frac{\epsilon^x_{\bm k}-\epsilon^z_{\bm k}}{i\hbar\omega_{\bm q}} \langle  L_{\bm q,\bm k}^y\rangle, \\
\langle  L_{\bm q}^z\rangle &= \sum_{\bm k} \frac{2C_{\bm q}\dot{u}_{\bm q}^\lambda (f_{\bm k}^y-f_{\bm k}^x)}{\hbar^2\omega_{\bm q}^2 - (\epsilon^x_{\bm k}-\epsilon^y_{\bm k})^2}, \label{Lzsol}  \\
\langle  Q^{xy}_{\bm q,\bm k} \rangle &= \frac{\epsilon^y_{\bm k}-\epsilon^x_{\bm k}}{i\hbar\omega_{\bm q}} \langle  L_{\bm q,\bm k}^z\rangle, 
\end{align}
by neglecting the fast-oscillating terms.
The full form of the equations is given as
\begin{align}
&\hbar\frac{d \langle L_{\bm q,\bm k}^y \rangle }{dt} \approx (\epsilon_{\bm k}^x-\epsilon_{\bm k}^z) \langle Q_{\bm q,{\bm k}}^{zx} \rangle - 2i \lambda C_{\bm q}u_{\bm q}^\lambda (f_{\bm k}^z - f_{\bm k}^x ),  \label{eq:Ly} \\
& \hbar \frac{d \langle L_{\bm q,\bm k}^z \rangle }{dt} \approx  (\epsilon_{\bm k}^y-\epsilon_{\bm k}^x) \langle Q_{\bm q,\bm k}^{xy} \rangle - 2 C_{\bm q}u_{\bm q}^\lambda (f_{\bm k}^y - f_{\bm k}^x) , \label{eq:Lz} \\
&\hbar \frac{d \langle Q_{\bm q,\bm k}^{jx} \rangle}{dt} \approx  - \sum_{k} \epsilon_{xjk} (\epsilon_{\bm k}^x-\epsilon_{\bm k}^j) \langle L_{\bm q,\bm k}^{k} \rangle, \label{eq:Q}
\\
\label{eq:Lx}
&\hbar \frac{d \langle  L_{\bm Q=0,\bm k}^x \rangle }{dt} \approx  (\epsilon_{\bm k}^z-\epsilon_{\bm k}^y) \langle  Q_{\bm Q=0,\bm k}^{yz} \rangle -i \sum_{\bm q'} C_{\bm q'}\Bigg\{u_{\bm q'}^+  \langle  Q^{-}_{-\bm q',\bm k} \rangle - u_{\bm q'}^- \langle  Q^{+}_{-\bm q',\bm k} \rangle  \Bigg\}, \\
\label{eq:Q_yz}
&\hbar\frac{d \langle  Q_{\bm Q=0,\bm k}^{yz} \rangle}{dt} \approx - (\epsilon_{\bm k}^z-\epsilon_{\bm k}^y)\langle  L_{\bm Q=0,\bm k}^{x} \rangle + \sum_{\bm q'} C_{\bm q'} \Bigg\{ u_{\bm q'}^+ \langle  L^{-}_{-\bm q',\bm k} \rangle + u_{\bm q'}^- \langle  L^{+}_{-\bm q',\bm k} \rangle \Bigg\},
\end{align}
where we assumed $|\bm q'|\ll|\bm k|$ and $\langle c^\dag_{\bm k,\pm}c_{\bm k-\bm q',x}\rangle \approx \langle c^\dag_{\bm k+\bm q',\pm}c_{\bm k}\rangle$.
Firstly, Eqs.~\eqref{eq:Ly} and \eqref{eq:Lz} indicate that the lattice displacement directly generates the orbital moments of $\langle L^y_{\bm q,\bm k}\rangle$ and $\langle L^z_{\bm q,\bm k}\rangle$ via electron-phonon coupling when there is an imbalance among $f^x$, $f^y$, and $f^z$.
For instance, a lattice displacement along the $z$ direction, $u_z$, activates the mixing between the $p_z$ and $p_x$ orbitals through the quadrupole operator $Q^{zx}$, which acts on $\ket{p_z}$ as $\exp\left(-iQ^{zx}\vartheta\right)\ket{p_z}=\cos\vartheta\ket{p_z}+i\sin\vartheta \ket{p_x}$, and $\langle L^y\rangle $ is generated.
Subsequently, when the $p_x$ and $p_z$ orbitals have different energies, their time evolution produces a relative phase factor of $\exp\left(-i(\epsilon^x-\epsilon^z) \Delta t\right)$, resulting in a finite expectation value of $Q^{zx}$, which corresponds to Eq.~\eqref{eq:Q}. 
Note that since $\vartheta$ temporally oscillates due to the phonon frequency, the resulting $\langle Q^{zx}\rangle$ remains after the rapid oscillation associated with the electron energies is averaged out.
Finally, the $u_y$ component of the chiral lattice displacement acts through the electron-phonon coupling associated with $Q^{xy}$. Acting on $\ket{p_x}$, it generates $\cos\varphi \ket{p_x} + i\sin\varphi\ket{p_y}$. Combined with the already generated $\ket{p_z}$ component, this entire process produces a mixing between $\ket{p_z}$ and $\ket{p_y}$, which effectively corresponds to a finite $\langle L^x\rangle$ and $\langle Q^{yz}\rangle$ as indicated in Eq.~\eqref{eq:Lx} and \eqref{eq:Q_yz}. 
In this way, the chiral phonon first induces an oscillatory interorbital motion and then converts it, through successive quadrupolar couplings, into the orbital accumulation.
When we consider the displacement of $u_{\bm q}^\lambda$, the second term of Eq.~\eqref{eq:Lx} becomes time-independent because of $\langle  Q_{-\bm q,\bm k}^-\rangle \propto u_{-\bm q}^{\bar\lambda}$, and we obtain a closed form for the equation of motion of the orbital angular momentum as
\begin{align}\label{eq:L_x_2}
&\hbar^2 \frac{d^2\langle  L_{\bm Q=0,\bm k}^x\rangle}{dt^2} \approx -(\epsilon^z_{\bm k}-\epsilon^y_{\bm k})^2 \langle  L_{\bm Q=0,\bm k}^x\rangle +(\epsilon^z_{\bm k}-\epsilon^y_{\bm k}) \sum_{\bm q'} C_{\bm q'} \Bigg\{ u_{\bm q'}^+ \langle  L^+_{-\bm q',\bm k} \rangle + u_{\bm q'}^- \langle  L^-_{-\bm q',\bm k} \rangle \Bigg\}.
\end{align}
Finally, we obtain the time-independent orbital accumulation as
\begin{align}
&\langle  L_{\bm Q=0,\bm k}^x\rangle \approx \frac{1}{\epsilon_{\bm k}^z-\epsilon_{\bm k}^y} \sum_{\bm q'} C_{\bm q'} \Bigg\{ u_{\bm q'}^+ \langle  L^+_{-\bm q',\bm k} \rangle + u_{\bm q'}^- \langle  L^-_{-\bm q',\bm k} \rangle \Bigg\} ,
\end{align}
with the neglect of the fast-oscillating terms.
By substituting Eqs.~\eqref{Lysol} and \eqref{Lzsol}, we reproduce the result based on the Berry phase given in Eqs.~(19) and (21) in the main text.

\section{Additional model result for varying $t_\pi$}
\begin{figure}
    \centering
    \includegraphics[width=0.6\linewidth]{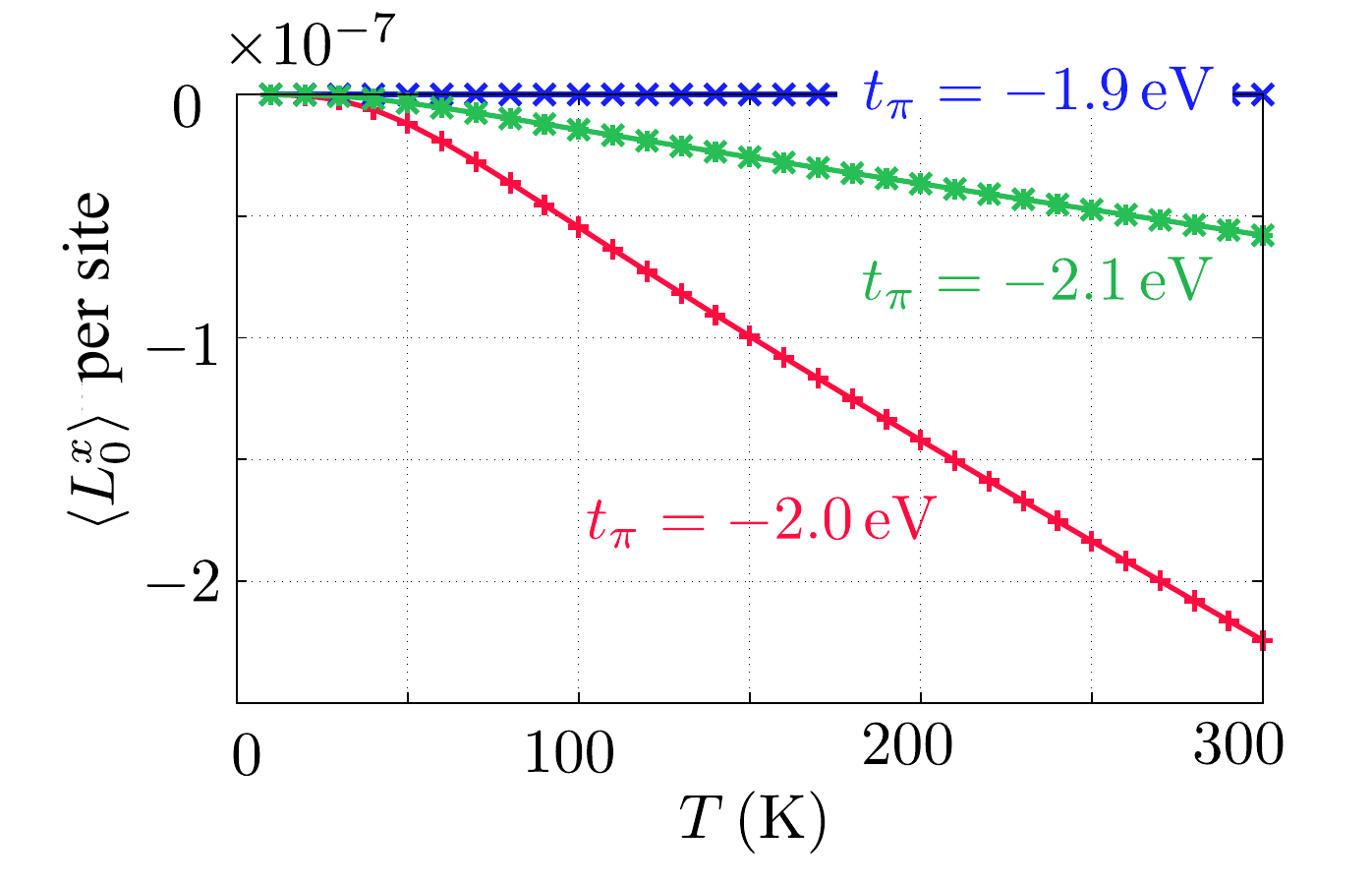}
    \caption{$t_{\pi}$ is changed as $-1.9,\, -2.0,\,-2.1\,{\rm eV}$, respectively, with $t_{\sigma} = -1.9\,{\rm eV}$. The Fermi energy for $t_\pi = -2.0\,{\rm eV}$ is set to be $\epsilon_F=-1.5\,{\rm eV}$, and for other $t_\pi$, the Fermi energy is adjusted to have the same number of electrons.}
    \label{figS:tpi}
\end{figure}

For completeness, we also show in Fig.~\ref{figS:tpi} the temperature dependence of $\langle L_x^0\rangle$ for several values of $t_\pi$ in the $p$-orbital model discussed in the main text.
Fig.~\ref{figS:tpi} shows the orbital accumulation when $t_\pi$ is varied while maintaining the electron density fixed.
Since the strength of the electron-phonon coupling is proportional to $t_\sigma - t_\pi$, the magnitude of the orbital accumulation is larger for $t_\pi = -2.0\,{\rm eV}$ than for $t_\pi = -1.9 \,{\rm eV}$.
However, as indicated by the result for $t_\pi = -2.1 \,{\rm eV}$, the orbital accumulation is suppressed when $t_\pi$ is further lowered.
This is because the denominator of ${\cal A}_{\bm k}$ includes information about the bandwidth.

\section{Numerical calculation for ${\rm Sr}_2{\rm RuO}_4$}

$\text{Sr}_2\text{RuO}_4$ is modeled on a square-lattice tight-binding model in which the electronic states are dominated by three $t_{2g}$ orbitals, $d_{xy}$, $d_{yz}$, and $d_{zx}$. The hopping symmetry among these orbitals is analogous to that of $p$ orbitals, with the correspondences $d_{yz} \leftrightarrow p_x$, $d_{zx} \leftrightarrow p_y$, and $d_{xy} \leftrightarrow p_z$.
Following Ref.~\cite{Cobo2016,Philippe2021}, the electronic dispersions are given by $\epsilon_{\bm k}^{yz} = -2t_{2}\cos k_x d - 2t_{1}\cos k_y d - \mu$, $\epsilon_{\bm k}^{zx} = -2t_{1}\cos k_x d - 2t_{2}\cos k_y d - \mu$, and $\epsilon_{\bm k}^{xy} = - 2t_3(\cos k_x d + \cos k_y d) - 4t_4 \cos k_x d \cos k_y d - 2t_5(\cos 2k_xd + \cos 2k_y d) - \mu$, where $t_1=88\,{\rm meV}$, $t_2=9\,{\rm meV}$, and $t_3=80\,{\rm meV}$ are the nearest-neighbor hopping amplitudes, $t_4=40\,{\rm meV}$ is the next-nearest-neighbor hopping amplitude, $t_5=5\,{\rm meV}$ is the third-nearest-neighbor hopping amplitude, and $\mu = 109\,{\rm meV}$ is the chemical potential.
For these hopping parameters, $t_1$ and $t_3$ are nearly equal since both correspond to $t_\pi$-like hopping processes.
In our derivations of the electron-phonon interaction, we approximately set $t_1=t_3$ and neglect next-nearest-neighbor and third-nearest-neighbor hopping for simplicity.
As a result, the electron-phonon interaction in $\text{Sr}_2\text{RuO}_4$ reduces to the same form as Eq.~(10) in the main text under the replacements $t_1=t_3 \to t_\pi$ and $t_2 \to t_\sigma$.

\begin{figure}[ht]
    \centering
    \includegraphics[width=0.5\linewidth]{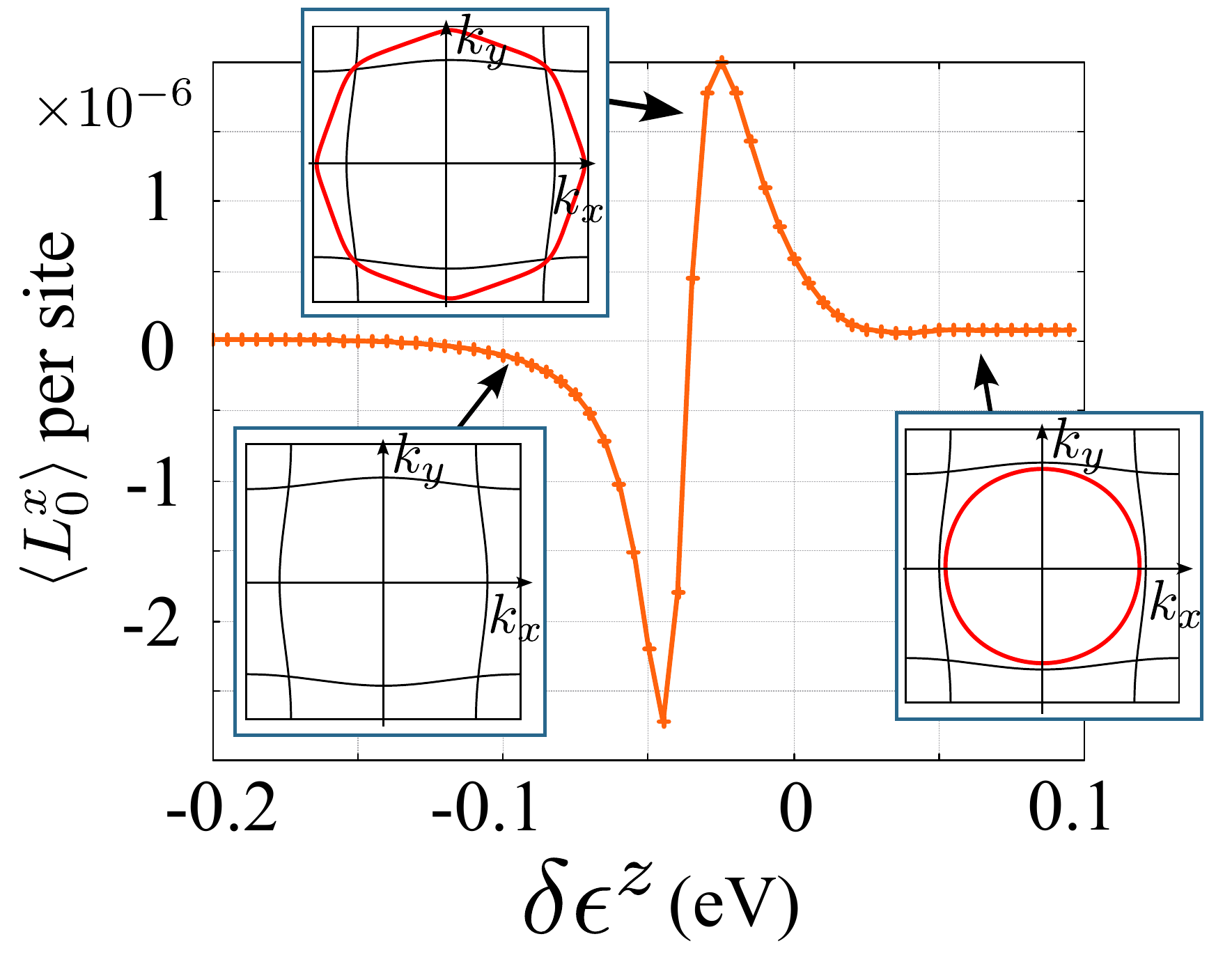}
    \caption{
$\delta\epsilon^z$ dependence of orbital accumulation at $200\,{\rm K}$. Inset: Fermi surfaces for the three $t_{2g}$ orbitals. The red circle represents the $d_{xy}$ orbital, and the black lines represent the $d_{yz}$ and $d_{zx}$ orbitals. The outer square indicates the Brillouin zone.}
    \label{fig:Sr2RuO4_2}
\end{figure}

Based on these settings, we performed numerical calculations following Eq.~(24).
Figure 4(b) in the main text shows the calculated orbital accumulation as a function of temperature.
Since the long-wavelength approximation $k \ll 1/d$ is not applicable to $\text{Sr}_2\text{RuO}_4$, we replaced $q_x$ in the coefficient $C_q$ with the full expression $\sin(k_x + q_x)d - \sin k_x d$.
Next, Fig.~\ref{fig:Sr2RuO4_2} shows the change in orbital accumulation when the energy of the $d_{xy}$ orbital is artificially shifted by $\delta \epsilon^z$, together with the corresponding Fermi surfaces in the insets. 
Consistent with the results in the main text, we find that the orbital accumulation is enhanced when degeneracy occurs near the Fermi energy.

\bibliography{reference}